\documentclass[conference,compsoc,a4paper,10pt,times]{IEEEtran}
\ifCLASSOPTIONcompsoc
  \usepackage[nocompress]{cite}
\else
  \usepackage{cite}
\fi
\ifCLASSINFOpdf
\else
\fi
\usepackage{tikz}
\usepackage{amsmath}

\usepackage{comment}
\usepackage{amsmath,amssymb,amsfonts}
\usepackage{algorithmic}
\usepackage{algorithm}
\usepackage{graphicx}
\usepackage{textcomp}
\usepackage{xcolor}
\usepackage{listings}
\usepackage[numbers,sort&compress]{natbib} %
\usepackage{indentfirst}

\usepackage{caption}
\usepackage{tcolorbox}
\usepackage[newfloat]{minted}
\usepackage{xcolor}
\usepackage{framed}
\usepackage{bbding}
\newcommand{\blzs}{\textsuperscript{$\blacklozenge$}}
\usepackage[strict]{changepage}

\usepackage{framed}

\usepackage{booktabs}
\usepackage{multirow}
\usepackage{multicol}
\usepackage{xcolor}
\usepackage{adjustbox}
\usepackage{pifont} % <-- Required for checkmarks and crosses
\usepackage{array} % multiline columns
\usepackage{url}
\usepackage{colortbl}

\usepackage{tikz}
\usepackage{forest}
\usetikzlibrary{shapes.geometric} % LATEX and plain TEX when using TikZ
\usetikzlibrary{shadows} %shadow effects for figures

\usepackage{arydshln}

\usepackage[normalem]{ulem}

\tcbset{
  insight/.style={
    colback=green!40!yellow!10!white,    % light cyan/teal background
    colframe=black,           % black border
    boxrule=0.8pt,            % border thickness
    arc=3pt,                  % rounded corners
    left=4pt, right=4pt, top=4pt, bottom=4pt, % padding
  }
}

\usepackage{rotating}
\usepackage{float}
\usepackage{hyperref}
\hypersetup{colorlinks=true,linkcolor={blue},citecolor={blue},urlcolor={black}}
\definecolor{Gray}{gray}{0.85}
\newcolumntype{a}{>{\columncolor{Gray}}p}
\newcolumntype{z}{>{\columncolor{white}}p}

\newcommand{\cmark}{\ding{51}} % Checkmark ✓
\newcommand{\xmark}{\ding{55}} % Cross ✗
\newcommand{\uarch}{$\mu$Arch~} % command to generate μArch without having to use the real syntax
\newcommand{\uarchn}{$\mu$Arch} % command to generate μArch without having to use the real syntax
\newcommand*\mycircle[1]{%
  \begin{tikzpicture}[baseline=-1pt]
    \node[draw,circle,inner sep=1pt] {#1};
  \end{tikzpicture}}

\newcommand*\mysquare[1]{%
  \begin{tikzpicture}[baseline=-1pt]
    \node[draw,rectangle,inner sep=2pt] {#1};
  \end{tikzpicture}}

\newcommand*\mytriangle[1]{%
  \begin{tikzpicture}[baseline=-1pt]
    \node[draw,regular polygon, regular polygon sides=3,inner sep=0.1pt] {#1};
  \end{tikzpicture}}

\newcommand{\ignore}[1]{}

\newcounter{insightct} % Defines a new counter named 'mycounter', reset by 'section'
\newcounter{limitationct}
\newcounter{tradeoffct}

\newcommand{\insightcount}[1]{}
\newcommand{\limitationcount}[1]{}
\newcommand{\tradeoffcount}[1]{}

\renewcommand{\insightcount}{\arabic{insightct}\stepcounter{insightct}} % Defines how the counter is displayed (e.g., 1.1, 1.2)
\renewcommand{\limitationcount}{\arabic{limitationct}\stepcounter{limitationct}}
\renewcommand{\tradeoffcount}{\arabic{tradeoffct}\stepcounter{tradeoffct}}

\begin{document}
\title{SoK: ARCUS: On the Efficiency and Efficacy of Hardware Fuzzing}

%\author{\IEEEauthorblockN{Michael Shell}
	%\IEEEauthorblockA{Georgia Institute of Technology\\
%		someemail@somedomain.com}
%	\and
%	\IEEEauthorblockN{Homer Simpson}
%	\IEEEauthorblockA{Twentieth Century Fox\\
	%	homer@thesimpsons.com}
%	\and
	%\IEEEauthorblockN{James Kirk\\ and Montgomery Scott}
	%\IEEEauthorblockA{Starfleet Academy\\
	%	someemail@somedomain.com}}
	


\author{\IEEEauthorblockN{Alenkruth Krishnan Murali\IEEEauthorrefmark{1}, Raghul Saravanan\IEEEauthorrefmark{2}, \\Sai Manoj P D\IEEEauthorrefmark{2}, and Ashish Venkat\IEEEauthorrefmark{3}}\\
\IEEEauthorblockA{\textit{\IEEEauthorrefmark{1}Department of Electrical and Computer Engineering,}
\textit{University of Virginia, Charlottesville, VA, USA}\\
\textit{\IEEEauthorrefmark{2}Department of Electrical and Computer Engineering,}
\textit{George Mason University, Fairfax, VA, USA}
\\
\textit{\IEEEauthorrefmark{3}Department of Computer Science,}
\textit{University of Virginia, Charlottesville, VA, USA}
\\
\{alenkruth, venkat\}@virginia .edu, \{rsaravan, spudukot\}@gmu.edu
}
}

%Telephone: (800) 555--1212, Fax: (888) 555--1212}


\maketitle

\begin{abstract}

This work presents a comprehensive analysis of contemporary hardware fuzzing techniques applied across three major abstraction layers: Instruction Set Architecture (ISA), microarchitecture, and Register-Transfer Level (RTL). Our study examines key factors including input stimulus quality, mutation strategies, feedback mechanisms, target platforms, reference models, and achieved coverage. We find challenges, goals, and design trade-offs vary significantly across abstraction layers. We further identify several unmet needs in current hardware fuzzing practices, such as intelligent input generation, reliable and scalable golden reference models, expressive feedback channels, and cross-layer integration. Building on these insights, we outline future research directions, including hybrid fuzzing frameworks, AI-assisted test generation, scalable reference models, standardized evaluation metrics and benchmarks, and human-in-the-loop automation for guided exploration and analysis. Together, they aim to unlock efficient, reliable, and comprehensive hardware verification solutions.

%--for instance, RTL fuzzers benefit from deep visibility but rely on limited coverage metrics, while ISA-level fuzzers focus on instruction semantics and compliance.

\end{abstract}


% \fi
\IEEEpeerreviewmaketitle

% With Moore's Law approaching its limitations \cite{}, a pressing need emerges for groundbreaking innovations in hardware design. These innovations are essential to furnish state-of-the-art, application-specific solutions that can keep pace with evolving technological requirements \cite{}.

\section{Introduction} \label{sec:intro}
The growing complexity of modern processors has created a fertile ground for bugs and vulnerabilities~\cite{Artenstein'17, Liu'17, Lipp'18, Kocher'18}, causing costly failures and security exposures~\cite{Intel'19,Mitre}. Hardware verification is a critical step in hardware design flows to ensure system integrity and yet, existing techniques struggle to scale with increasingly complex architectures~\cite{Dessoky'19, Sarangi'06, Wagner'07, Chen'11, Mukherjee'15}. 
To address these challenges, fuzzing~\cite{fuzzing-seminal-unix-utilities} has been recently adopted to the hardware world to assist in verification~\cite{ossfuzz, domas-sandsifter, skipscan, uisfuzz, examiner, iscanu, idev, liblisa, mishegos, n-version-disassambly, riscvuzz, revizor-oleksenko, medusa-moghimi, speechminer-xiao, osiris-weber, scamv, observationrefinement, Saravanan'24b}, allowing seamless scaling to complex architectures, while also exploring previously unseen paths and uncovering novel bugs.

\begin{figure}[htb!]
%\vspace{-1em}
  \centering
  \includegraphics[width=\linewidth]{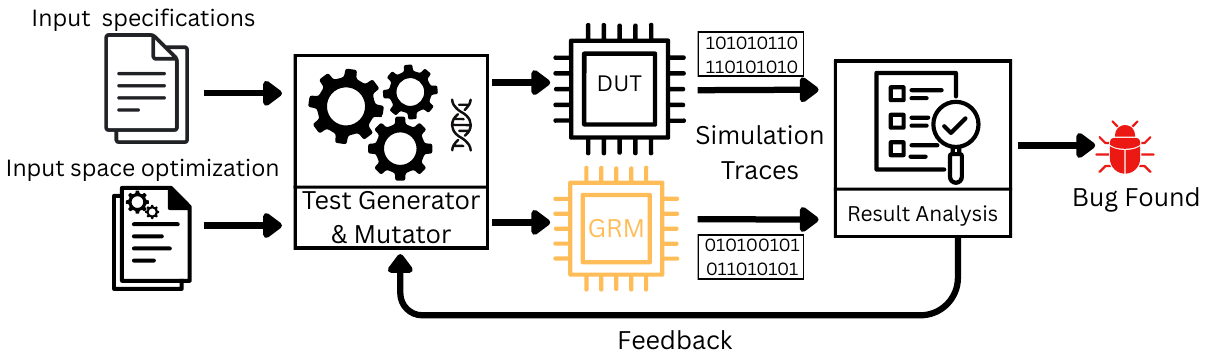} 
  %\vspace{-1em}
  \caption{Overview of hardware fuzzing: Inputs are generated based on specifications, tests are run on the target and Golden Reference Model (GRM) if available, feedback is collected from the target to guide the fuzzer and identify bugs. }
  \vspace*{-0.12in}
  \label{fig:fuzzfig}
  %\vspace{-1em}
\end{figure}

\textbf{Hardware Fuzzing. } Figure~\ref{fig:fuzzfig} illustrates the operation of a typical hardware fuzzer that systematically feeds the Design Under Test (DUT), such as a processor with random inputs to activate unexplored parts of the design and potentially trigger bugs. The inputs are then iteratively mutated based on the feedback from the DUT until the desired coverage is reached, similarly to software fuzzers. 
%American Fuzzy Lop (AFL)-style mutation operations (Appendix \ref{sec:afl} gives a non-exhaustive list of mutations used) are preferred. 
Typically, hardware fuzzers detect bugs by comparing the execution traces of a DUT with that of a Golden Reference Model (GRM), a fast high-level model of the DUT. GRMs do not mirror the low-level details of the DUT but maintain the programmer-visible state accurately. 
% A common GRM for RISC-V CPUs is SPIKE~\cite{SPIKE}, the RISC-V ISA simulator. \textcolor{red}{Previous sentence can be commented. }
Appendix~\ref{sec:appendix-SW-to-HW} provides a mapping of ideas and terms between software and hardware fuzzing.

%While some hardware fuzzers target designs at the gate-level, we consider them out of scope for this study since they could be visualized as an extension of RTL fuzzing. 
%\textit{Fuzzing at different abstraction layers presents unique layer-specific challenges, making a direct comparison of fuzzers targeting different layers difficult.} Moreover, the lack of a comprehensive study of existing hardware fuzzers further limits their understanding. Thus, a unified and systematic evaluation is therefore essential to provide holistic insights and guide future hardware fuzzing research.

\begin{figure}
  \centering
  \includegraphics[width=1\linewidth]{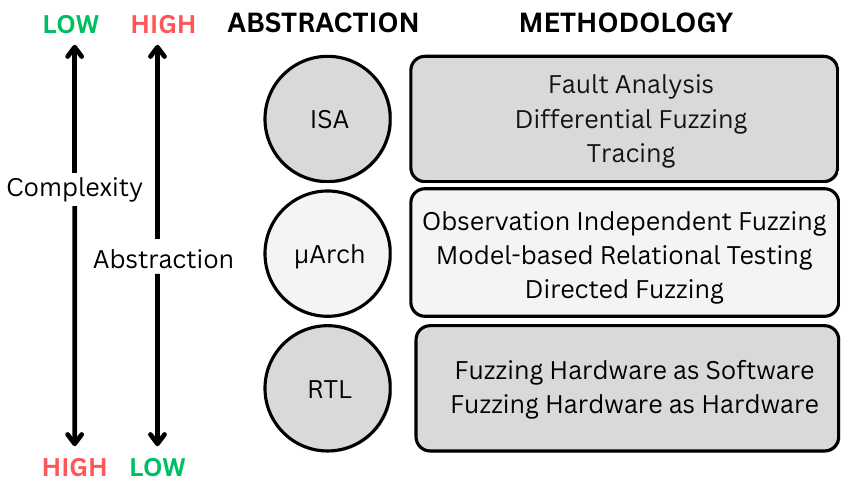}
  \caption{Classification of Hardware Fuzzers based on the level of abstraction on the Hardware stack.}
  \label{fig:classification}
\end{figure}

We find hardware fuzzing targets are diverse---black-box commodity CPUs, instruction decoders, disassemblers, emulators, hardware intellectual property (IP), and white-box register-transfer level (RTL) sources and synthesizers, spanning multiple abstraction layers that each present their own unique set of challenges. This divergence in fuzzing goals across different targets and abstraction layers influences critical design choices, including input generation, coverage extraction, and GRM selection, necessitating tailored methodologies. For example, fuzzers can use feedback from the white-box DUT to improve fuzzing efficiency akin to software fuzzing~\cite{feedback-in-fuzzing}, but extracting feedback from a black-box commodity CPU is difficult, requiring fuzzers to rely on imprecise software-visible components such as performance counters for feedback. Moreover, fuzzing goals and complexity differ by the target abstraction layer as shown in Figure~\ref{fig:classification}. Instruction Set Architecture (ISA)-level fuzzers seek to identify \textit{deviations from the ISA specification} like undocumented instructions or decoder flaws in\textit{ commodity black-box CPUs (post-silicon)}. Whereas, microarchitectural fuzzers typically aim at \textit{uncovering unspecified yet exploitable microarchitectural behavior} like novel side channels in \textit{commodity black-box CPUs (post-silicon)}. Finally, RTL fuzzing aims to identify \textit{design implementation bugs} in \textit{white-box RTL designs (i.e pre-silicon)}. This diversity results in a complex and fragmented landscape that is challenging to navigate without systematic studies that taxonomize and classify hardware fuzzing methodologies across different targets and abstraction layers.

Prior efforts that compare and evaluate fuzzers~\cite{encarsia, saravanan2024Odyssey, kevin} are limited in their scope in that they only examine fuzzers targeting white-box RTL designs. Encarsia~\cite{encarsia} studies four fuzzers, Cascade~\cite{cascade-razavi}, DiFuzzRTL~\cite{Hur'21-difuzzrtl}, ProcessorFuzz~\cite{Canakci'23-processorfuzz}, and RFuzz~\cite{Laeufer'18}, providing eight insights and five recommendations. The Fuzz-Odyssey~\cite{saravanan2024Odyssey} surveys twelve white-box RTL-level fuzzers and identifies challenges associated with choosing the right coverage metrics, and Golden Reference Models (GRMs). While valuable, existing works do not sufficiently systematize the breadth of hardware fuzzers that have been proposed.

\begin{table}[t]
\centering
\scriptsize
\renewcommand{\arraystretch}{1.2}
\begin{tabular}{p{2cm}p{6cm}}
\hline
\textbf{Dimension} & \textbf{Description} \\
\hline
\textbf{Input Stimulus} & The type of test input provided to the DUT. The fuzzer mutates these inputs during the campaign. \\
\textbf{Fuzzing Methodology} & The overall testing strategy adopted by the tool, e.g., differential fuzzing or fault analysis. \\
\textbf{Fuzzing Target} & The hardware design or intellectual property (IP) module under test. \\
\textbf{Black/White-Box Target} & Whether the DUT is treated as a black box. Target observability impacts coverage metrics and feedback channels. \\
\textbf{Coverage~\& Feedback} & Whether runtime feedback from the DUT is used to guide the fuzzer during exploration. \\
\textbf{Fuzzing Algorithm} & The mutation strategy for generating new input stimuli, including search space filtering and optimizations for rapid test generation. \\
\textbf{GRM} & Whether the fuzzer relies on a GRM to detect deviations or to generate feedback. \\
\textbf{Type of GRM} & The specific type of reference model employed (e.g., functional simulator, ISA model). \\
\textbf{Automation} & Whether the framework automates the analysis of outputs and reporting of findings. \\
\textbf{Bugs Found} & The categories of bugs uncovered, which may vary depending on tool design, target, and abstraction level. \\
\bottomrule
\end{tabular}
\caption{Analysis Dimensions for Hardware Fuzzers.}
\label{tab:dimensions}
\end{table}

In this work, we introduce \textbf{\textit{ARCUS}}~\footnote{\textit{Arcus} (Latin), meaning arch or curve, symbolizing a structural link or bridge. It reflects our goal of connecting and unifying diverse research in hardware fuzzing through a coherent, comprehensive study.}, the first systematic analysis of contemporary hardware fuzzers targeting diverse hardware design targets and verification goals.  We develop a two-level taxonomy -- the first level classifies each approach by the abstraction layer at which they operate, i.e., ISA (post-silicon), \uarch(post-silicon), and RTL(pre-silicon), and the second level groups works within each abstraction layer into subclasses by the fuzzing approach/methodology inherent to that layer (as shown in Figure~\ref{fig:classification}).  We then analyze each class and subclass across multiple common dimensions (summarized in Table~\ref{tab:dimensions}) that are key to evaluating the efficacy of any given hardware fuzzer.  These dimensions reflect fundamental aspects of hardware fuzzing rather than tool-specific implementation details, ensuring both robustness and extensibility of the classification.   
% \textcolor{red}{Why is ISA and \uarch post-silicon?}

Each analysis dimension captures a core property that influences the efficacy, scalability, and applicability of a hardware fuzzer. For example, the nature of the input stimulus and mutation strategy shapes the exploration of the design space, while the availability of coverage feedback and golden reference models governs the precision with which bugs can be isolated.  Similarly, distinguishing between black-box and white-box targets allows us to assess the feasibility of applying fuzzers across different stages of the hardware design life cycle. % \textcolor{red}{To me keeping this \papercount sounds more like a limitation -- rather avoid it, unless we have a specific need to emphasize in introduction.}
These dimensions also sufficiently explore the diversity offered by the hardware fuzzing strategies we explore as summarized in Table~\ref{tab:comparison}.  Collectively, our classification and multi-dimensional analysis strike a balance between breadth and depth: they provide a sufficiently granular basis for distinguishing among existing tools, yet remain abstract enough to accommodate emerging methodologies and future innovations in hardware fuzzing.

%The divergence in the objectives and challenges at different abstraction levels influences critical design choices, including input generation, coverage extraction, and the selection of reference models, thereby necessitating tailored methodologies for each abstraction layer. Such diversity introduces a varied set of dimensions and trade-offs that shape the design and overall effectiveness of the fuzzers.

%We develop a novel taxonomy of hardware fuzzers within and across the different abstraction levels, drawing key insights regarding their efficacy across various metrics.  

This work also identifies several unmet requirements in existing hardware fuzzing approaches, including the need for expressive and reliable feedback, intelligent input generation, scalable and trustworthy golden reference models, and stronger integration between abstraction levels. Building on these insights, we outline future research directions, such as hybrid fuzzing pipelines that bridge multiple target abstraction layers and the adoption of AI-driven input generation and mutation techniques to enhance adaptability and efficiency.

In summary, we make the following major contributions:
\begin{itemize}\itemsep0em
\vspace{-0.5em}
    \item We conduct the first comprehensive analysis of hardware fuzzers across the ISA, $\mu$Arch (post-silicon) and RTL (pre-silicon) layers.  
    %\vspace{-0.1em}
    \item We introduce a two-level taxonomy that classifies fuzzers by abstraction layer and fuzzing methodology.  
    %\vspace{-0.1em}
    \item We define core analysis dimensions (e.g., input stimulus, fuzzing algorithm, coverage feedback, GRMs) and use them to systematically standardize and compare existing approaches.  
    %\vspace{-0.1em}
    \item We identify key shortcomings, including limited feedback mechanisms, weak input generation, reliance on GRMs, and poor cross-layer integration.  
    %\vspace{-0.1em}
    \item We outline future research opportunities to improve precision, scalability, adaptability, and efficiency.  
\end{itemize}

\begin{table*}[htbp]
    \renewcommand{\arraystretch}{1}
    \caption{Comparison of Hardware Fuzzing Approaches}
    \label{tab:comparison}
    \small
    \begin{adjustbox}{max width=\textwidth}
    %\rowcolors{2}{gray!25}{white}
    \begin{tabular}
        % Type | Methodology | Tool | Input | Target | Coverage | Fuzzing Algo. | GRM | GRM Type | Manual Analysis | Bugs
        {p{0.7cm}a{2cm}p{1.8cm}p{1cm}p{1.2cm}p{1cm}p{5cm}p{1cm}p{1.5cm}p{1.5cm}p{2.5cm}}
        \toprule
        \toprule
        \textbf{Type} & \textbf{Methodology} & \textbf{Tool} & \textbf{Target} & \textbf{Input} & \textbf{Feedback} & \textbf{Fuzzing Algorithm} & \textbf{GRM?} & \textbf{GRM Type} & \textbf{Manual Analysis?} & \textbf{Types of Bugs Identified} \\
        \midrule
        
        \multirow[b]{10}{*}[-40pt]{\rotatebox{90}{ISA Fuzzing}} & DF \& FA & Sandsifter\blzs\cite{domas-sandsifter} & \mysquare{C} & X &  \cmark & Random & \cmark & ISA\textsubscript{Emu}, ISA\textsubscript{Doc}, Dis & \cmark & UInst, Emu, Dec\\
        %\cdashline{2-12}[2pt/2pt]
        & DF \& FA & UISFuzz\cite{uisfuzz}  & \mysquare{C} & X & \cmark & Constrained random & \cmark & ISA\textsubscript{Emu}, ISA\textsubscript{Doc}, Dis & \textit{recheck} & UInst, Emu, Dec\\
        %\cdashline{2-12}[2pt/2pt]
        & DF \& FA & SkipScan\cite{skipscan} & \mysquare{C} & X & \cmark & Constrained random & \cmark & ISA\textsubscript{Emu}, ISA\textsubscript{Doc}, Dis & \xmark & UInst, Emu, Dec \\
        %\cdashline{2-12}[2pt/2pt]
        & DF \& FA & iScanU\blzs\cite{iscanu} & \mysquare{C} & A \& R &  \xmark & Exhuastive search & \cmark  & ISA\textsubscript{Emu} & \cmark  & UInst, Emu \\
        %\cdashline{2-12}[2pt/2pt]
        & DF & Examiner\blzs\cite{examiner} & \mysquare{C} \mysquare{E} & A & \xmark & SAT solver using ISA semantics. & \cmark & ISA\textsubscript{Emu}, HW & \cmark & UInst, Dec, Emu, SEM\textsubscript{HW}, SEM\textsubscript{SW}, Inc \\
        %\cdashline{2-12}[2pt/2pt]
        & DF & iDEV\cite{idev} & \mysquare{C} \mysquare{E} & A & \xmark & Constrained by ISA encodings & \cmark\textsuperscript{N} &  ISA\textsubscript{Emu}, HW & \cmark & UInst, Emu, Dec, SEM\textsubscript{HW}, SEM\textsubscript{SW}, Inc \\ 
        %\cdashline{2-12}[2pt/2pt]
        & DF & RISCVuzz\blzs\cite{riscvuzz} & \mysquare{C(s)} \mysquare{E} & R & \xmark & Constrained by ISA encodings & \cmark\textsuperscript{N} & HW, ISA\textsubscript{Emu} & \cmark & UInst, Emu, Ghostwrite, UFault \\ 
        %\cdashline{2-12}[2pt/2pt]
        & DF & N-version Disassembly \cite{n-version-disassambly}& \mysquare{SD} & X & \cmark & Random + GRM validation & \cmark & HW & \cmark & Dec\\
        %\cdashline{2-12}[2pt/2pt]
        & DF & Mishegos\blzs\cite{mishegos} & \mysquare{SD} & X & \xmark & Random + \textit{Sliding strategy} & \xmark & & \cmark & Dec \\
        %\cdashline{2-12}[2pt/2pt]
        & Tracing & libLISA\blzs\cite{liblisa} & \mysquare{C} \mysquare{E} & X & \cmark & Random + Feedback driven refinement & \cmark & Dis &  \cmark & UInst, Emu, Dec, UFault, SEM\textsubscript{HW}, SEM\textsubscript{SW}, Inc\\
        
        \midrule
        
        \multirow[b]{6}{*}[-30pt]{\rotatebox{90}{$\mu$Arch Fuzzing}} & MRT & Revizor\blzs\cite{revizor-oleksenko} & \mysquare{C} & X & \cmark\textsuperscript{pt} &  Random + \textit{Pattern coverage} & \cmark\textsuperscript{ExCon} & ISA\textsubscript{Emu} & \cmark & Spectre V1 \& V4 \\
        %\cdashline{2-12}[2pt/2pt]
        & OMRT & ScamV1\blzs~\cite{scamv} & \mysquare{C} & A & \cmark\textsuperscript{symb} & Random or model-based & \xmark & & \cmark & Cache channels, UFault, ARM ISA-model bugs.  \\
        %\cdashline{2-12}[2pt/2pt]
        & OMRT & P. Buiras et al.\blzs~\cite{observationrefinement} (ScamV2) & \mysquare{C} & A & \cmark\textsuperscript{symb} & Random or model-based + Feedback refinement & \xmark & & \cmark & SISCLOAK \\         
        %\cdashline{2-12}[2pt/2pt]
        & OIF & Medusa\blzs\cite{medusa-moghimi} & \mysquare{C} & X & \xmark & Mutated Meltdown & \xmark & & \xmark & Medusa \& Meltdown.\\
        %\cdashline{2-12}[2pt/2pt]
        & OIF & SpeechMiner \cite{speechminer-xiao} & \mysquare{C} & X & \xmark & Constrained random + Feedback & \xmark & & \cmark & Spectre \& Meltdown  \\
        %\cdashline{2-12}[2pt/2pt]
        & DiF & ABSynthe \cite{absynthe-gras} & \mysquare{C} & X \& A & \cmark & Guided mutation + Feedback & \xmark & & \xmark & Contention-based channels \\
        & OIF & Osiris\blzs\cite{osiris-weber} & \mysquare{C} & X & \xmark & Exhaustive search + GRM validation & \xmark & & \xmark & Timing-based channels  \\     
        
        \midrule

        \multirow[b]{13}{*}[-30pt]{\rotatebox{90}{RTL Fuzzing}} & FHaS\textsuperscript{f} & RFuzz\blzs\cite{Laeufer'18} & C, IP & B & \cmark \textsuperscript{M} & AFL mutation  & \xmark &  & \xmark &  \\
        %\cdashline{2-12}[2pt/2pt]
        & FHaS\textsuperscript{f} \& DF & DiFuzzRTL\blzs\cite{Hur'21-difuzzrtl} & C &  R &  \cmark \textsuperscript{R} & AFL mutation  & \cmark & ISA\textsubscript{Emu}  & \cmark & Inc, Func  \\
        %\cdashline{2-12}[2pt/2pt]
        & FHaS \& SFH & HyperFuzzing\blzs\cite{Muduli'20-hyperfuzzing} &  SoC & B \& P & \cmark \textsuperscript{N,T} & AFL mutation   & \cmark & SP, C  & \xmark & FT \\
        %\cdashline{2-12}[2pt/2pt]
        & FHaS \& SFH & Trippel et. al.\blzs\cite{Trippel'22-hw-as-sw} & IP & B & \cmark \textsuperscript{} & AFL Mutation & \xmark & C  & \cmark & FT  \\
        %\cdashline{2-12}[2pt/2pt]
        & FHaS & Cascade\blzs\cite{cascade-razavi} & C & R & \xmark & Constrained Random + ISA\textsubscript{Emu} validation & \xmark & ISA\textsubscript{Emu} & \xmark & Func, Synth \\
        %\cdashline{2-12}[2pt/2pt]
        & FHaS & FeedbackFuzz\cite{feedbackfuzz} & C & R & \cmark\textsuperscript{Dis} & Constrained Random + ISA\textsubscript{Emu} validation & \xmark & ISA\textsubscript{Emu} & \textit{LLM} & Func, Synth \\
        %\cdashline{2-12}[2pt/2pt]
        & FHaS & SIGFuzz\blzs\cite{sigfuzz} & C & R & \xmark & Random generation + Guided mutation + Feedback & \xmark & & \xmark & Func, Timing-based channels \\
        %\cdashline{2-12}[2pt/2pt]
        & FHaH & SoCFuzzer\cite{Hossain'23-socfuzzer} & SoC & R & \cmark \textsuperscript{CoSP} & AFL mutation + cost function  & \cmark & VD  & \cmark & Func \\ 
        %\cdashline{2-12}[2pt/2pt]
        & FHaH & FormalFuzzer\cite{Farzana'19-formalfuzzer} & SoC & R \& AS & \cmark \textsuperscript{CoSP} & AFL mutation  & \cmark & VD  & \cmark & Func \\
        %\cdashline{2-12}[2pt/2pt]
        & FHaH \& DF & LogicFuzzer\cite{logicfuzzer-jose-renau} & C & R & \xmark &
        ISA Tests (\cite{riscv-tests}, \cite{riscv-dv}) & \cmark & ISA\textsubscript{Emu} & \cmark & Func \\
        %\cdashline{2-12}[2pt/2pt]
        & FHaH \& DF & WhisperFuzz\cite{borkar2024whisperfuzz} & C & R & \cmark \textsuperscript{B,T,F,E,L} & AFL mutation  & \cmark & ISA\textsubscript{Emu}  & \cmark & Inc, Timing-based channels  \\ 
        %\cdashline{2-12}[2pt/2pt]
        & FHaH \& DF & TheHuzz\cite{Kande'22-thehuzz} & C & R & \cmark \textsuperscript{B,T,F,E,L} & AFL mutation & \cmark & ISA\textsubscript{Emu}  & \cmark & Inc, Func  \\ 
        %\cdashline{2-12}[2pt/2pt]
        & FHaH \& DF & HyPFuzz\cite{hybrid-hypfuzz} & C & R \& AS & \cmark \textsuperscript{B,T,F,E,L} & AFL mutation + formal engine  & \cmark & ISA\textsubscript{Emu}  & \cmark & Inc, Func  \\ 
         & FHaH \& DF & ProcessorFuzz\blzs\cite{Canakci'23-processorfuzz} & C & R & \cmark \textsuperscript{CSR} & AFL mutation  & \cmark & ISA\textsubscript{Emu}  & \cmark & Inc, Func  \\
         & FHaH \& DF & PSOFuzz\cite{psofuzz} & C & R & \cmark \textsuperscript{B} & AFL mutation + \textbf{P}article \textbf{S}warm \textbf{O}pt & \cmark & ISA\textsubscript{Emu}  & \cmark & Inc, Func  \\ 
          & FHaH \& DF & MABFuzz\cite{mabfuzz} & C & R & \cmark \textsuperscript{B} & AFL mutation + \textbf{M}ulti-\textbf{A}rmed \textbf{B}andit & \cmark & ISA\textsubscript{Emu}  & \cmark & Inc, Func  \\ 
          & FHaH \& DF & ChatFuzz\cite{chatfuzz} & C & R & \cmark \textsuperscript{B} & AFL mutation + LLM & \cmark & ISA\textsubscript{Emu}  & \cmark & Inc, Func  \\ 
            & FHaH \& DF & HFL\cite{hfl}, RLFuzz\cite{rlfuzz} & C & R & \cmark \textsuperscript{B,T,F,E,L} & AFL mutation + RL + LSTM & \cmark & ISA\textsubscript{Emu}  & \cmark & Inc, Func  \\ 
           & FHaH \& DF & GraphFuzz\cite{graphfuzz}& IP & B & \cmark \textsuperscript{GT} & AFL mutation & \cmark & ISA\textsubscript{Emu}  & \cmark & Inc, Func  \\ 
            & FHaH \& DF & SynFuzz\cite{synfuzz}& C, IP & B & \cmark \textsuperscript{T,E} & AFL mutation & \cmark & R, Syn & \cmark & Synth  \\

        %\cdashline{2-12}[2pt/2pt]
       
          & FHaH \& DF & GenHuzz\blzs\cite{genhuzz} & C & R & \cmark \textsuperscript{B,T,F,E,L} & AFL mutation + LLM & \cmark & ISA\textsubscript{Emu}  & \cmark & Inc, Func  \\

        \bottomrule
        \bottomrule
        \multicolumn{11}{c}{\textbf{Legend}} \\
        \midrule

        \multicolumn{2}{r}{\multirow[c]{3}{*}{\textbf{Methodology}}} & \multicolumn{9}{l}{FA : Fault Analysis~~~DF : Differential Fuzzing~~~OIF : Observation-Independent Fuzzing~~~MRT : Model-based Relational Testing~~~OMRT : Observation-Model RT} \\
        
        \multicolumn{2}{c}{} & \multicolumn{3}{l}{SFH : Software Fuzzers for Hardware} & \multicolumn{6}{l}{FHaS : Fuzzing Hardware as Software~~~FHaH : Fuzzing Hardware as Hardware~~~DiF : Directed Fuzzing} \\

        \multicolumn{2}{c}{} & \multicolumn{6}{l}{FHaS\textsuperscript{f} : Fuzzing Hardware as Software with an FPGA Emulation Option} \\
        
        \multicolumn{2}{r}{\textbf{Tool}} & \multicolumn{5}{l}{\blzs : open-source} \\    

        \multicolumn{2}{r}{\textbf{Target}} & \multicolumn{2}{l}{\mysquare{B} : Black-box} & \multicolumn{2}{l}{C : Processor} & \multicolumn{1}{l}{SD : Software Decoders~~~E : Emulators} & \multicolumn{2}{l}{IP : Standalone IP} & \multicolumn{2}{l}{SoC : System on Chip} \\

        \multicolumn{2}{r}{\textbf{Input}} & \multicolumn{2}{l}{X: x86~~~AS : RTL Assertions} & \multicolumn{2}{l}{A : ARM~~~R : RISC-V} & \multicolumn{3}{l}{B : Bits~~~P : HyperPLTL Security Properties} \\
        
        \multicolumn{2}{r}{\multirow[c]{2}{*}{\textbf{Feedback}}} & \multicolumn{1}{l}{\xmark : No coverage} & \multicolumn{2}{l}{\cmark : Previous instructions} & \multicolumn{2}{l}{\cmark \textsuperscript{B,T,F,E,L} : Branch, Toggle, FSM, Expression, Line} & \multicolumn{2}{l}{\cmark\textsuperscript{pt} : Pattern coverage} & \multicolumn{2}{l}{\cmark\textsuperscript{Dis} : Speculative execution} \\
        
        \multicolumn{2}{c}{} & \multicolumn{1}{l}{\cmark \textsuperscript{M} : Multiplexer} & \multicolumn{2}{l}{\cmark \textsuperscript{CSR} : CSR Transitions} & \multicolumn{2}{l}{\cmark\textsuperscript{symb} : Symbolic execution~~~\cmark \textsuperscript{R} : Register Coverage} & \multicolumn{2}{l}{\cmark \textsuperscript{N} : NoC Transactions} & \multicolumn{2}{l}{\cmark \textsuperscript{CoSP} : Security policy cost func.} \\
        
        \multicolumn{2}{r}{\textbf{GRM}} & \multicolumn{1}{l}{\cmark : One GRM}& \multicolumn{2}{l}{\cmark\textsuperscript{N} : $>$1 GRM} & \multicolumn{2}{l}{\cmark\textsuperscript{ExCon} : Execution Contract} \\
        
        \multicolumn{2}{r}{\multirow[c]{2}{*}{\textbf{GRM Type}}} & \multicolumn{1}{l}{C : Crashes} & \multicolumn{2}{l}{ISA\textsubscript{Emu} : ISA Emulators} & \multicolumn{2}{l}{HW : commodity CPU~~~ISA\textsubscript{Doc} : ISA Documentation} & \multicolumn{4}{l}{SP : Security Properties~~~VD : Vulnerability Database} \\
        
        \multicolumn{2}{c}{} & \multicolumn{2}{l}{Dis : Disassembler} & \multicolumn{2}{l}{R : RTL} & \multicolumn{2}{l}{Syn : Golden Synthesis}\\

        \multicolumn{2}{r}{\multirow[c]{2}{*}{\textbf{Types of Bugs}}} & \multicolumn{3}{l}{Emu : Emulator Bugs~~~FT : Firmware Tests} & \multicolumn{4}{l}{UInst : Undisclosed Instructions~~~Dec : SW Decoder/Disassembler Bugs} & \multicolumn{2}{l}{UFault : Undefined Faults} \\

        \multicolumn{2}{c}{} & \multicolumn{2}{l}{Func : Functional Bugs} & \multicolumn{2}{l}{Synth : Synthesizer bugs} & \multicolumn{1}{l}{Sem\textsubscript{X} : Semantic Deviation in X} & \multicolumn{3}{l}{Inc : Inconsistency between HW \& SW} \\
     
        \bottomrule
        \bottomrule
        %% papers not in the table but would need to be in the analysis
        %% 1. SpecFuzz (?), 2. ABSynthe 3. Encarsia 
        %% 4. Type A - On a Consistency Testing Model and Strategy for Revealing RISC Processor’s Dark Instructions and Vulnerabilities
        %% 5. Examiner Pro 
    \end{tabular}
    \end{adjustbox}
%%\end{table*}
    %% \vspace{5mm}

\end{table*}

\vspace*{-0.05in}
\section{ISA-Level Fuzzing} \label{sec:typea}

%\subsection{Overview}
\label{sec:typea-overview}

% The Instruction Set Architecture (ISA) defines the contract between software and hardware, specifying the architecturally visible behavior of instructions. ISA-level fuzzing targets black-box, post-silicon commodity CPUs to identify deviations from ISA compliance, such as undocumented instructions, incorrect decoding, or non-conformant execution. Crucially, ISA fuzzing differs from RTL and microarchitectural fuzzing in three key ways: (1) it operates on \textit{black-box hardware} without internal visibility (unlike white-box RTL fuzzing), (2) it targets \textit{violations of explicitly specified} architectural behavior (unlike microarchitectural fuzzing, which exploits unspecified implementation details), and (3) it occurs \textit{post-silicon} when design changes are prohibitively expensive. These ISA deviations can lead to security vulnerabilities, incorrect behavior, or exploitable undefined states in real-world systems~\cite{riscvuzz}. Additionally, ISA fuzzers try to identify inconsistencies between different processor generations implementing the same ISA, and semantic deviations from specification in processors, emulators, and disassemblers. Historically, commodity processors have been considered to be trusted black boxes. Likewise, assemblers and emulators have been trusted to faithfully model the execution semantics as specified in manufacturer documentation. The ISA fuzzers examined in this section challenge this assumption. 

The Instruction Set Architecture (ISA) defines the contract between software and hardware, specifying the architecturally visible behavior of instructions. ISA-level fuzzing targets black-box, post-silicon commodity CPUs to identify deviations from ISA compliance, such as undocumented instructions, incorrect decoding, or non-conformant execution. ISA fuzzing differs from RTL and microarchitectural fuzzing in two key ways: (1) it operates on black-box hardware without internal visibility (unlike white-box RTL fuzzing), and (2) it targets violations of explicitly specified architectural behavior (unlike microarchitectural fuzzing, which exploits unspecified implementation details). These deviations can lead to security vulnerabilities, incorrect behavior, or exploitable undefined states~\cite{riscvuzz}. ISA fuzzers also identify inconsistencies between processor generations implementing the same ISA and semantic deviations in processors, emulators, and disassemblers, challenging the historical assumption that commodity processors, assemblers, and emulators are trusted black boxes that faithfully implement ISA specifications.

% , and (3) it occurs post-silicon when design changes are prohibitively expensive. 
% These deviations can lead to security vulnerabilities, incorrect behavior, or exploitable undefined states~\cite{riscvuzz}. ISA fuzzers also identify inconsistencies between processor generations implementing the same ISA and semantic deviations in processors, emulators, and disassemblers---challenging the historical assumption that commodity processors, assemblers, and emulators are trusted black boxes that faithfully implement ISA specifications.

% \textcolor{red}{One thing to add here is ``what exactly does ISA fuzzing look like?" In other words, what is basis for considering a work as ISA-level fuzzing. You can state that ISA level fuzzers consider or focus on fuzzing ISA inputs and deal with xyz... This is also one of the concerns raised in previous reviews. You can do same for other 2 types of fuzzers -- I know you have details later, but adding a sentence or two would be helpful.}

% \textcolor{red}{In Table~\ref{tab:comparison}, why do you have X, A, R as input. Should we make it as ISA? I guess when you refer to X, A, R -- it means ISA of those architectures, right? for \uarch you have specified ARM as input type instead of A -- standradize it. Fixed. X, A, and R indicate x86, RISC-V and ARM ISAs } 

As shown in Table~\ref{tab:comparison}, all ISA fuzzers are similar in the types of inputs they generate, and their target being black-boxes, and in the requirement of a golden reference model. Additionally, these fuzzers consider their targets to be black-boxes, thereby limiting their ability to use prior design knowledge to guide input generation. Once the input stimulus is generated through fuzzing, it is executed on their target(s) and, if present, on golden reference model(s). Based on how the fuzzing methodology is carried out, we classify ISA-level fuzzers into three subclasses as shown in Figure~\ref{fig:classification}. 
In this Section, we expand on the subclasses and the other minor differences present in the fuzzers that are subtle or present across members of all subclasses, and discuss their implications.

%\textcolor{red}{Why most? Why not all? If, only most, what is basis for putting those not under "most" into this bin? We mainly discuss the differences inside the respective dimension's section. adding the outliers here will bloat text }
%\textcolor{red}{Reference to fig 2 should come earlier. This is sort of too late to introduce fig 2. Done}
%\textcolor{red}{What are subclasses? Are you referring to subclass within the ISA/RTL/uarch or these 3 are subclasses?}

\vspace*{-0.05in}
\subsection{Fuzzing Targets and Input Stimulus} \label{sec:typea-target-dut-feedback}
%% coverage and feedback would be a point of discussion for all classes. Particularly class B. Class A and C assume black box targets limiting the scope for coverage guided generation.
%As mentioned earlier, ISA fuzzers aim to identify bugs and flaws in a diverse set of DUTs, namely, commodity CPUs, emulators like QEMU, and disassemblers/decoders. \textbf{
The commonality between the works studied are that all of their targets are black-boxes, i.e., there is no prior knowledge of their targets.  LibLISA~\cite{liblisa} and iDEV~\cite{idev} identify semantic deviation between manufacturer documentation and emulators, disassemblers, and even multiple generations of the CPUs. Since extracting semantics of an instruction requires fuzzing the input space, iDEV~\cite{idev} and LibLISA~\cite{liblisa} also identify decoder bugs in CPUs and uncover undocumented/misdocumented instructions. Mishegos~\cite{mishegos} and N-version disassembly~\cite{n-version-disassambly} are strictly software emulator fuzzing tools. We include them in our study primarily because their fuzzing approach, despite targeting software implementations, is applicable to CPU fuzzing. Likewise, SkipScan, a predominantly hardware fuzzer, also tests a wide variety of disassemblers and detects several implementation bugs in addition to undocumented instructions in the CPUs.
% \textcolor{red}{I would probably put part of this paragraph at end of section, as this is more like exception rather than generic case.}

\ignore{\textcolor{red}{The following insight is not well reflected in this text.}

\begin{tcolorbox}[insight]
    \textbf{Insight:} Limited/Lack of feedback from black-box DUTs necessitate the generation of huge test corpus and the adoption differential-fuzzing and/or fault analysis to identify bugs.
\end{tcolorbox}
}

% Fuzzing a block-box target is difficult, particularly because receiving feedback from the tests is either impossible or feedback is very limited. Lack of feedback means that the fuzzers will have to generate a huge test corpus to effectively fuzz the design. Additionally, the lack of feedback necessitates differential-fuzzing or fault-based analysis to identify bugs. We discuss input generation and methodologies in the following sections.  

ISA fuzzers typically generate instruction streams as inputs to the DUT and further extract runtime information using those streams to guide subsequent fuzzing campaigns. Given their black-box targets, the fuzzers need to intelligently collect runtime information from the target if possible, or perform other optimizations to the input search space. In this section we focus how inputs are generated in the studied fuzzers.

\textbf{Generation Strategies: } The na\"ive approach to generate instructions would be to exhaustively search the instruction space~\cite{iscanu}. Modern ISAs support large sets of opcodes, addressing modes, and extensions making an exhaustive search intractable. Therefore, fuzzers optimize the generation in two ways -- structured generation and random generation. To impose structure in instruction generation, one could constrict the search space by only checking a sub-set of the prefixes or opcodes. This way, the inputs are going to be limited to a specific region of the search space. An extension to this would be to skip exhaustively covering immediate and displacement fields and only testing corner case values like -1, 0, 1, and so on~\cite{skipscan, idev, riscvuzz}. Another structured approach would be to extract the structure (constraints) imposed on instructions by the ISA from the specification, and use that as the back-bone to generate new instructions for fuzzing~\cite{examiner}. Structured generation is of low complexity, but it comes at the cost of limiting exploration to find corner-case bugs.

The second optimization is random generation~\cite{domas-sandsifter, uisfuzz, mishegos, n-version-disassambly}. Randomized instruction generation would allow generation of diverse test-cases that could improve the quality of testing.  However, complete randomness can hurt fuzzing efficiency as there is no control over the generated tests and the low probability of a random test identifying a deeply-embedded bug. To that end, most ISA fuzzers check the randomly generated instructions on a GRM~\cite{n-version-disassambly, mishegos} before running on the target. When the goal of fuzzing is to identify undocumented instructions, complete random instruction generation of valid instructions is better than fully structured generation as random generation is neither constrained by the ISA specification nor limited to a subset of the input space.
\begin{table}
\centering
\footnotesize
\resizebox{\linewidth}{!}{
\begin{tabular}{|l|p{2cm}|p{2cm}|p{2cm}|}
\hline
\textbf{Tool} & \textbf{\# Total Insts.} & \textbf{\# Undoc. Insts.} & \textbf{Test Efficiency} \\
\hline
Sandsifter~\cite{domas-sandsifter} & 936,606,326 & 3,961,284 &  1x\\% 343m \\
\hline
UISFuzz~\cite{uisfuzz} & 131,270,176 & 2,157,079 &  5.57x\\ %60m \\
\hline
SkipScan~\cite{skipscan} & 242,102,430 & 3,961,284 & 4x\\ %$\sim$90m \\
\hline
\end{tabular}
}
\caption{Comparison of input generation. Data from SkipScan~(\cite{skipscan}\S4). Test efficiency is the ratio of valid instructions to the total instructions tested, normalized to Sandsifter.} % Runtime calculated when tools generated instructions with a 1-Byte prefix on an Intel i7-8700k processor @4.20GHz.}
\label{tab:typea-input-effectiveness-undocumentedinsts}
\vspace*{-0.15in}
\end{table}

\textbf{Discussion: }Sandsifter \cite{domas-sandsifter} generates random inputs guided by fault analysis, and UISFuzz \cite{uisfuzz} generates random inputs just like Sandsifter but adds an instruction format analysis step that identifies and skips displacement and immediate fields. SkipScan improves on UISFuzz by not blindly skipping immediate and displacement fields and by identifying and using rules governing the prefix fields. UISFuzz's aggressive approach to skip immediate and displacement fields affects the instruction length of legal instructions, skips several reserved instructions and also skips several Mod R/M bytes. Table~\ref{tab:typea-input-effectiveness-undocumentedinsts}, we observe that the rule-driven structure enforced on random inputs by SkipScan helps it search less instructions than Sandsifter and UISFuzz while identifying the same number of undocumented instructions as Sandsifter.

% \textcolor{red}{I guess random input mean random ISA, right? Yes, random instructions.}

% From the comparison data, we found that UISFuzz is more efficient than Skipscan, however, UISFuzz ignores the test of some reserved instructions. UISFuzz [3] shrinks the number of immediate and displacement operands from 28∗n to 1, which is aggressive and affects the instruction length in some legal instructions. Thus, UISFuzz fails to find consider-able undocumented instructions, the number of undocumented instructions detected by UISFuzz is 2,157,079, which is smaller than 3,961,284 (number before rechecking). Although UISFuzz only miss few opcodes of undocumented instruction (0F C7), it skips lots of ModR/M bytes of instructions 0F 18 and 0F {1A-1F}.

\begin{tcolorbox}[insight] \label{insight:1}
    \textbf{Insight~\insightcount:} 
    Randomized generation with imposed structure (e.g., SkipScan, Mishegos) effectively identifies undocumented instructions, disassembler bugs, and decoder bugs by diversifying tests while limiting the search space to areas of interest.
    %Randomized generation with some imposed structure (like SkipScan, and Mishegos) to generate valid instructions is effective in identifying undocumented instructions, disassembler bugs, and decoder bugs because, the randomization helps diversify the tests while the structuring helps limit the search space to areas of interest.
\end{tcolorbox}

\vspace*{-0.05in}
\subsection{Fuzzing Methodology} \label{sec:typea-methodology}

Fuzzing black-box DUTs is difficult. In order to overcome the difficulties introduced by the lack of feedback, ISA fuzzers adopt certain methods based on which we classify them further. The three subclasses are: (a) Fault Analysis, (b) Differential Fuzzing, and (c) Tracing. 
%  as discussed above. Tracing-based approaches, in contrast, collect runtime information—such as instruction traces, architectural states, or execution paths—to detect behavioral anomalies. Note that these methodologies are complementary and are often combined to increase fuzzing effectiveness.

\vspace*{-0.05in}
\subsubsection{Fault Analysis} \label{sec:typea-methodology-faultanalysis}
Fault Analysis is a coarse grained technique aimed at identifying violations in expected behavior by monitoring system responses such as exceptions, crashes, or incorrect state transitions during test execution. These methods often rely on observing signals such as segmentation faults, page faults, or assertion failures to detect anomalies in processor behavior~\cite{domas-sandsifter, uisfuzz, skipscan, iscanu}. Fault Analysis is well-suited for black-box fuzzing to identify undocumented instructions as it does not require detailed specification about the target or semantic knowledge of instruction behavior while also being a reliable source of feedback to guide fuzzing. %Faults do not incur significant overhead as all fuzzing targets tend to have some fault handling mechanism set up, and the test only needs to check for the occurrence of the fault when trying to find undocumented instructions. 

% Sandsifter~\cite{domas-sandsifter}, UISFuzz~\cite{uisfuzz}, Skipscan~\cite{skipscan}, and IScanU~\cite{iscanu} are Fault Analysis-based ISA-level fuzzing techniques. They identify undocumented instructions by observing the faults (or the lack thereof) raised by the processor when trying to execute them and compare them with a GRM. 

The instructions that are decoded by the target but are absent in the ISA documentation or cannot be decoded by a disassembler/GRM are deemed to be undocumented instructions. Observing the fault type helps differentiate between actual undocumented instructions and instructions restricted to a specific privilege ring. This is important because executing privileged instructions from a lower privilege mode would throw a fault code different from an undefined instruction. Nonetheless, fault analysis in fuzzing is \textit{limited by 1) its blindness to bugs that do not manifest as observable exceptions and 2) lack of precision in fault handling due to hypervisors, emulators, and other unknown processor behavior}.

\subsubsection{Differential Fuzzing} \label{sec:typea-methodology-differentialfuzzing}

Differential Fuzzing involves comparing the outputs of the DUT against a GRM for the same input; any deviation in the results, i.e., register states, memory values, or decoded outputs, gets flagged as a bug. Since differential fuzzing leverages behavioral equivalence, it is especially powerful when no formal ISA specification is available, making it ideal for testing proprietary, legacy, or undocumented hardware.  Moreover, differential fuzzing allows for systematic exploration of semantic correctness across instruction encodings, addressing a broad class of subtle bugs that fault analysis would miss.  However, it also hinges heavily on the correctness and coverage of the GRM.

\textbf{Further Classification: } We classify differential fuzzing at the ISA level into three primary classes based on the type of Golden Reference Model (GRM) used for comparison. %T Class 2, in contrast, employs software models—such as simulators, disassemblers, or emulators—as the reference. These are often more accessible and easier to instrument, making them popular in fuzzing environments where rapid iteration is needed. Class 3 combines hardware and software references. By comparing against multiple types of references, it can also help isolate bugs originating from either the DUT or an unreliable GRM.
The first class leverages hardware as the GRM, wherein the test execution on the target is validated against a hardware implementation.
N-version Disassembly~\cite{n-version-disassambly} differentially fuzzes x86 instruction decoders in disassemblers with an x86 processor as an Oracle/GRM. It flags the disassemblers that disagree with the GRM are incorrect, and the higher the number of agreeing disassemblers, the greater the confidence in their results. 
%The paper also introduces a normalization mechanism to normalize the outputs from the target disassemblers.

The next class, in contrast, employs software models such as simulators, disassemblers, or emulators as the reference.  For example, Mishegos~\cite{mishegos} uses majority consensus among nine disassemblers as a proxy for an oracle to identify probable errors (\textit{N-Oracle Approach}). %Oversupporting discrepancy occurs when majority of the disassembler agree that an instruction is \textit{invalid}, whereas the converse applies for undersupporting. Missupporting is when majority of the disassemblers agree that the instruction's length or semantics differ consistently. 
Our final class combines hardware and software references. By comparing against multiple types of references, it can also help isolate bugs originating from either the DUT or an unreliable GRM.  iDEV~\cite{idev} is an example of such a fuzzer.  iDEV looks inconsistencies between -- (a) the CPU decoding an instruction, (b) disassemblers decoding an instruction, and (c) between the disassembler and the CPU decoding an instruction.  During execution of the generated test case, iDEV sets up state and monitors the signals generated. Examiner~\cite{examiner} extends iDEV with a better testing methodology, and by evaluating on a diverse set of ARM ISA variants and emulators. The differential testing engine in Examiner~\cite{examiner} stores the target CPU and reference emulator state (memory, registers, PC, status flags, exceptions) before and after test execution. The inequivalence of states indicates an inconsistency in either the CPU or the emulator or both. %RISCVuzz~\cite{riscvuzz} differentially fuzzes commercially available RISC-V CPU implementations and multiple versions of the RISCV QEMU~\cite{qemu} emulator. %In addition to finding undocumented instructions, the input sequences used in RISCVuzz also reveal microarchitectural bugs and inconsistencies related to address translation, decoding, and fault handling.
% \textcolor{red}{Breakdown into smaller paragraphs}

% When testing emulators, a reference CPU is used. 

% Finally,  RISCVuzz sets up state, runs the generated test sequence and collects the state after instruction execution. 

\begin{table}[!htb]
\centering
\resizebox{\linewidth}{!}{%
\begin{tabular}{|l|p{1.2cm}|p{1.5cm}|p{1.2cm}|p{1.7cm}|}
\hline
\textbf{Tool} & \textbf{IDev} \cite{idev} & \textbf{Examiner} \cite{examiner} & \textbf{IScanU} \cite{iscanu} & \textbf{RISCVuzz} \cite{riscvuzz} \\
\hline
Avg. SW time (minutes) & 566\textsuperscript{board} & 9\textsuperscript{virtual-cpu} & - & - \\
\hline
Avg. HW time (minutes) & 106 & 245\textsuperscript{embedded} & - & - \\
\hline
Avg. SW throughput (Insts./s) & 988  & 1675  & - & - \\
\hline 
Avg. HW throughput (Insts./s) & 7534  & 112 & 781K  & 18.1-59.2K  \\
\hline
\end{tabular}
}
\caption{Performance comparison of hardware and software references. Data collected from respective papers.}
\label{tab:typea-throughput}
\end{table}

%% discussion
\textbf{Discussion: } \textit{The key limitation of differential fuzzing lies in the trustworthiness of the golden reference(GRM): a faulty or incomplete model can lead to false positives or, worse, hide real bugs}. The trustworthiness of the GRM is especially critical when using software references, which may themselves be incomplete or behaviorally divergent from specification. Hardware-based references suffer from limited observability and difficulty in debugging mismatches. The combined GRM/N-oracle approaches, while more robust, increase the complexity of test harnesses and require better analysis methods. Additionally, discrepancies due to undefined or implementation-specific behaviors must be carefully filtered, often requiring manual intervention. 

Depending on their implementation, software references can sometimes also increase testing time due their reduced throughput when compared to hardware references as shown in Table~\ref{tab:typea-throughput}. Virtualized CPU emulation using hypervisors like in Examiner is much faster than complete board emulation like in IDev. Likewise, speed of execution on hardware is dependent on the capabilities of the hardware (embedded CPUs in Examiner are slower than general purpose CPUs in IDev and IScanU). A combined GRM approach can improve testing throughput by only checking the faster GRM before advancing to the next test input while comparison against the slower GRM results happens in the background.

\ignore{
\begin{tcolorbox}[insight]
    \textbf{Insight~\insightcount:} Differential fuzzing needs high-quality reference models and careful engineering to ensure reliable results.
\end{tcolorbox}
}

\subsubsection{Execution Observation/Tracing} \label{sec:typea-methodology-tracing}

Tracing-based approaches in ISA-level fuzzing involve capturing and analyzing execution traces of test instructions on a processor or simulator to detect anomalies, inconsistencies, or potential vulnerabilities. These methods execute instruction sequences and observe detailed run-time behaviors such as memory accesses, register updates, execution timing, or microarchitectural events. The traced data is then compared against expected behavior, either derived from a specification, a model, or another execution environment. This enables identification of side effects, undocumented features, or microarchitectural nuances that may not be evident through fault analysis or differential fuzzing alone.

% LibLISA~\cite{liblisa} implements an observer that runs on a virtual machine. The observer sets up the state, executes the test instruction on the CPU from user space, and then saves the state changes. Like iDEV~\cite{idev}, 
LibLISA uses tracing to identify deviations in semantics of the implemented instructions. It uses the Intel Reference manual~\cite{x86-64} as its oracle to identify semantic inconsistencies between the specification and commodity processors. LibLISA identifies semantic deviations between multiple CPUs, undocumented/unsupported instructions, and software observable architectural deviations in the implementation of common arithmetic instructions. Moreover, the semantics generated by LibLISA could be used to fingerprint CPUs and build a semantically accurate x86 model to emulate userspace binaries.

\textbf{Discussion: } \textit{Tracing provides deep visibility into the execution pipeline, enabling detection of subtle inconsistencies in the targets, although it introduces high overhead, both in terms of performance and data volume.} Additionally, accurately interpreting traces requires significant domain knowledge, and the approach could miss bugs that do not manifest in observable differences. Finally, like differential fuzzing, tracing also suffers when GRMs are unreliable.
% \textcolor{red}{No insight or limitation box here?}

\vspace*{-0.05in}
\subsection{Golden Reference Models} \label{sec:typea-grms}
Fuzzers targeting a black box can benefit from a trusted golden reference model (GRM). GRMs are meticulously crafted and validated models that serve as benchmarks or standards for the expected behavior of a hardware design. %These models are considered the authoritative representations against which the actual hardware implementation is compared during verification and testing processes. 
A GRM can help validate/identify/classify the observations made from executing the test cases on the target. In the case of ISA fuzzing, depending on the target type, the fuzzers use a software, hardware, or both as their GRMs. Sometimes the fuzzers also consult the ISA documentation as one of the references while analyzing a discrepancy between the target and GRM. Having a GRM enables differential fuzzing since the same input is provided to both the target and the GRM. 
Examiner~\cite{examiner}, Mishegos~\cite{mishegos}, N-version disassembly~\cite{n-version-disassambly}, libLISA~\cite{liblisa}, and iDEV~\cite{idev} take advantage of multiple reference models with the assumption that neither of them is trusted. This \textit{N-oracle approach} avoids the situation where a single erroneous GRM leads to false positives and negatives.

Despite careful curation of GRMs, false positives and false negatives could arise from time to time.  For example, IScanU~\cite{iscanu} uses a disassembler as its GRM and a majority of the instructions flagged as undocumented were due to GRM errors.  Another example is a bug associated with incorrect memory handling in a GRM RISC-V core~\cite{Tricheck}. It is also essential that the versions of the ISA implemented by the target and the GRM match. For CPUs that implement custom instructions or custom extensions (as allowed by RISC-V), identifying GRMs that faithfully implement custom instructions and extensions could be difficult. 

\ignore{\begin{tcolorbox}[insight]
    \textbf{Limitation~\limitationcount:} Difficulties in identifying reliable GRMs for CPUs is further exacerbated with increasing CPU heterogeneity, instructions, and ISA extensions.
\end{tcolorbox}}

%\subsubsection{Discussion on GRMs and Manual Analysis} \label{sec:typea-grms-disc}

\subsection{Manual Analysis} \label{sec:typea-manual-analysis}

The black box-nature of the targets coupled with unreliable GRMs exacerbate false positives and false negatives. Without sufficient context, such as register state or memory trace, it is difficult to understand the cause of a discrepancy. While fuzzers discussed in this paper generate prologue and epilogue segments as a part of the generated tests to set the architectural state before test execution and collect the state changes after test execution, analysis tools are still limited to software-visible changes in the target.

Thus, \textit{most fuzzing approaches require some manual analysis to identify the cause of discrepancies with their GRMs or the cause of the fault}. UISFuzz~\cite{uisfuzz} introduces the notion of \textit{recheck} wherein the faulting instructions are queried against a database of up-to-date errata and ISA documentation from the manufacturer~\cite{x86-64}. If the query hits the database, the faulting instruction is not stored for further manual analysis. Unless there is a special filtering mechanism like \textit{recheck} from UISFuzz, manual effort would be required to identify the root cause of the faults. 
% \textcolor{red}{As 2.3 and 2.4 focus on bug finding part, should they be part of 1 subsection? Not really, I think they are separate because it talks about two different problems in bug finding. GRM lead to false-positives and negatives. MAnual analysis handles with triaging the bug irrespective of reference}

\ignore{\begin{tcolorbox}[insight]
    \textbf{Limitation~\limitationcount:} Manually analyzing the traces to identify and reason about the root cause is tedious and not scalable.  
\end{tcolorbox}}

\subsection{Bugs Detected from ISA-level Fuzzing} \label{sec:typea-bugs}

As shown in the Table~\ref{tab:comparison}, ISA-level fuzzing approaches identify different classes of bugs ranging from undisclosed instructions, incorrect decoder/inconsistent instruction implementations in hardware and software, semantic deviation from specifications, and semantic deviation among CPUs and emulators implementing the same ISA specification.

\section{Microarchitecture-Level Fuzzing} \label{sec:typeb}
%\subsection{Overview} \label{sec:typeb-overview}
% Microarchitectural (\uarchn)~Fuzzing aims to uncover subtle flaws and vulnerabilities in the hardware’s internal logic and behavior. Like ISA fuzzing, \uarch fuzzing also involves systematically generating and executing instruction sequences on the target and observing the execution, albeit for exposing hidden issues arising from typically undocumented design decisions such as timing side channels, speculative execution bugs, or undocumented interactions between pipeline components. As modern processors increasingly adopt aggressive performance optimizations, they become susceptible to microarchitectural leaks and inconsistencies that traditional verification methods might overlook, highlighting the critical importance of fuzzing at this level.

Microarchitectural (\uarchn) fuzzing aims to uncover vulnerabilities in the hardware's \textit{unspecified} internal implementation details. Unlike ISA fuzzing, which targets violations of explicitly documented architectural contracts, and RTL fuzzing, which verifies compliance with functional design specifications, microarchitectural fuzzing deliberately exploits the gap between what is \textit{specified} (the ISA) and what is \textit{implemented} (the microarchitecture). Like ISA fuzzing, it targets black-box post-silicon commodity CPUs by systematically generating and executing instruction sequences, but the goal is fundamentally different: exposing hidden behaviors arising from undocumented design decisions such as timing side channels, speculative execution vulnerabilities, or unintended interactions between pipeline components. These microarchitectural behaviors are neither bugs nor specification violations in the traditional sense. They are \textit{legitimate implementation choices} that become security vulnerabilities when exploited. As modern processors adopt increasingly aggressive performance optimizations (e.g., deep speculation, complex caching hierarchies), they become susceptible to microarchitectural leaks and inconsistencies that neither ISA compliance testing nor traditional RTL verification can detect, highlighting the critical importance of fuzzing at this level.

% thus plays a critical role in ensuring the security and robustness of processor designs.

% These vulnerabilities arise from complex, undocumented implementation details that are not captured in architectural specifications. A common example would be speculative execution and the vulnerabilities associated with it. 

%Note that approaches in Type C, if scalable to large designs, are suited to discover microarchitectural vulnerabilities with the right test-cases. Unfortunately, Type C fuzzers are not scalable with design complexity, and access to the RTL code of most commercial processors is proprietary making it impossible for a third-party to verify.

As shown in Table~\ref{tab:comparison}, most \uarch fuzzers are similar in the types of inputs they generate, and in their target type. These fuzzers consider their targets to be black boxes, but they use publicly available microarchitectural information to setup channels to guide input generation between fuzzing rounds. Once the input stimulus is generated through fuzzing, it is executed on their target(s) and, if present, on GRMs. %We classify microarchitectural black-box fuzzing into two classes based on their fuzzing methodology as shown Figure~\ref{fig:classification}. In this section we expand on the subclasses and the other minor differences present in the fuzzers that are subtle or present across members of all subclasses, and discuss their implications.
% \textcolor{red}{Probably difference between ISA and \uarch cn be made a little more clearer.}

%Like the ISA-level fuzzing approaches, microarchitectural fuzzers have several commonalities. Before discussing the classification we will briefly explain the commonalities and discuss their implications for black-box fuzzing.

\vspace*{-0.05in}
\subsection{Fuzzing Target and Input Stimulus} \label{sec:typeb-dut-feedback-analysis}
\uarch fuzzers typically fuzz black-box CPUs using minimal information about the micrarchitecture, to observe potential vulnerable side-effects of the generated test sequence. Designers use known information about the target to craft intelligent channels that leak runtime information to be used to guide the fuzzer. For example, the data cache acts as a side-channel in Revizor~\cite{revizor-oleksenko}, Medusa~\cite{medusa-moghimi}, Speechminer~\cite{speechminer-xiao}, ScamV~\cite{scamv}, and ScamV2~\cite{observationrefinement}.  Osiris~\cite{osiris-weber} seeks to identify non-contention-based timing channels by strategically placing serializing instructions around the executed code and then leveraging performance counters to observe side effects. 

\textbf{Generation Strategies: }\uarch fuzzing leverages well-constructed/targeted series of low-level assembly instructions/programs that exercise some specific part of the microarchitecture and cause an attacker-visible side-effect that could be used to deduce sensitive information. Similar to ISA fuzzing, one could adopt a structured test program generation or random generation for \uarch fuzzing. In the structured approach, the generated test programs contain three parts, 1) A trigger for the targeted microarchitectural vulnerability (speculation primitive), 2) Set of operations that delay the self-resolution of the trigger (windowing gadget), 3) Set of operations that propagate the changes resulting from the trigger to the channel for observation (disclosure gadget). 

Random generation approaches in \uarch fuzzing involves two steps. \textit{Step 1:} Generate valid random instructions. \textit{Step 2:} Rearrange them into basic blocks following specific structure to a) trigger the target vulnerability and b) avoid causing faults due to out-of-bounds memory accesses. Fuzzers could also mutate pre-existing attacks based on target information.% to identify if the new target is vulnerable. 

Speechminer~\cite{speechminer-xiao} leverages a structured input generation approach while Revizor~\cite{revizor-oleksenko}, and Osiris~\cite{osiris-weber} use constrained random generation where they rearrange the generated inputs to fit a specific pattern. Revizor checks the generated program to make sure they stay within allocated pages and Osiris limits itself to not testing control-flow instructions while using only a subset of registers and operands. Revizor introduces \textit{pattern coverage}, wherein the randomly generated basic blocks that do not exhibit a specific pattern  (load-after-load, load-after-store) are removed at the end of a run and the ones confining to the pattern are further mutated for subsequent runs.  Medusa~\cite{medusa-moghimi}, mutates known versions of Meltdown-style attacks~\cite{Lipp'18} to identify new variants on commodity processors. Likewise, ABSynthe~\cite{absynthe-gras} evolves a test program to create the best-performing side-channel. Finally, ScamV~\cite{scamv,observationrefinement} has the option to generate random programs or use a model-driven guided program generator. ScamV then uses a SMT solver to generate pairs of test cases  (observationally-equivalent states) for the program with respect to the model under test while also maximizing execution path coverage and cache-line coverage.

\begin{table}
\centering
\caption{Time to find a vulnerability for \uarch fuzzers. Data collected from respective papers.}
\scriptsize
\resizebox{\linewidth}{!}{%
\begin{tabular}{|l|p{1.3cm}|p{1.4cm}|p{1.2cm}|}
\hline
\textbf{Tool} & \textbf{Revizor}~\cite{revizor-oleksenko} & \textbf{Medusa}\textsuperscript{1}~\cite{medusa-moghimi} & \textbf{Osiris}\textsuperscript{2}~\cite{osiris-weber} \\
\hline
Time to Spectre V4 & 73m25s & N/A & N/A \\ 
\hline
Time to Spectre V1 &  4m51s & N/A & N/A \\
\hline
Time to MDS & 5m35s & 26h & N/A \\
\hline
Time to LVI & 7m40s & N/A & N/A \\
\hline
Time to Meltdown/MDS & N/A & 26h & N/A\\
\hline
Time to timing channel & N/A & N/A & $\sim$21 days\\
\hline
\end{tabular}
}
\footnotesize \raggedright {\textsuperscript{1} - Medusa's runtime is summed across 3 processors (Intel i7-7700, i7-8650U, and i9-9900K) finding 5100 snippets showing MDS-style transient leakage. \\ \textsuperscript{2} - Osiris' total runtime to identify multiple instances of 4 unknown timing channels.}
\vspace*{-0.2in}
\label{tab:typeb-time-comparison}
\end{table}

%symbolic execution engine that maximizes the coverage of all possible execution paths and all cache lines. The programs are paired with input states. The programs are first augmented with annotations that indicate observations for each statement. Then, to generate test cases for the program (i.e., two observationally equivalent states), they synthesize a relation to characterize the observationally equivalent states for the program with respect to the model under test using a symbolic execution engine.

%\subsection{Discussion on input generation} \label{sec:typeb-input-gen-disc}
\textbf{Discussion:} Unlike ISA fuzzing, wherein a random search of the instruction space could yield a counterexample, \textit{the quality of the generated inputs play a critical role in the efficiency of the  \uarch fuzzers.}  Since test sequences could be several instructions long, it is impossible to manually search the entire search space to find counterexample. Thus, constrained random generation/structural generation or mutation is commonly adopted by \uarch fuzzers. Although manual instruction generation methods are not scalable, the automated/multiphase approaches are effective at generating good counterexamples that exercise the target microarchitectural modules. As shown in Table~\ref{tab:typeb-time-comparison}, Revizor's approach to generate inputs based on "pattern-coverage" in multiple phases is effective at identifying bugs quicker. Osiris and Medusa generate a huge test corpus without feedback which leads to a huge runtime and ineffective test cases. %It should be noted that \uarch fuzzers like ISA fuzzers suffer from the mechanisms used to reduce the search space limiting the reach of test cases.

\begin{tcolorbox}[insight] \label{insight:2}
    \textbf{Insight~\insightcount:} 
    \uarchn itectural test generation complexity scales with bug complexity. Simple randomization and exclusion strategies that work for ISA fuzzing are ineffective for  \uarch fuzzing.
    % Test generation complexity increases with the complexity of targeted bugs. Randomization and exclusion strategies are ineffective for \uarch fuzzing.
\end{tcolorbox}

\vspace*{-0.05in}
\subsection{Fuzzing Methodology} \label{sec:typeb-methodology}

Black-box microarchitectural fuzzing can be broadly classified into three classes based on their fuzzing methodologies, i.e., how they explore the processor’s behavior and validate correctness: (a) Observation-Independent Fuzzing, (b) Model-based Relational Testing, and (c) Directed Fuzzing. These methodologies seek to overcome constraints placed by lack of internal visibility in black-box designs.

\vspace*{-0.05in}
\subsubsection{Observation-Independent Fuzzing} \label{sec:typeb-methodology-fuzztesting}

Observation-independent fuzzing approaches execute a large directed-test corpus generated through the structured approach on the target and analyze execution anomalies such as page faults, traps, or timing variations without relying on a GRM. The test cases are \textit{not} modified based on the observations and re-executed on the target. The key steps are: 1) set up microarchitectural state and a side-channel 2) execute well-crafted tests with specific triggers on the target and encode into the side-channel 3) observe the side-effects through the side-channel. Step 2 can be multi-phased where either the trigger and leakage are performed as distinct steps or the test is run with and without the trigger across two fuzzing campaigns. Step 3 can sometimes also include probing software visible performance counters to filter false-positives. 

Medusa~\cite{medusa-moghimi}, Speechminer~\cite{speechminer-xiao}, and Osiris~\cite{osiris-weber} fall into this category. Medusa uses an automated approach wherein information is leaked transiently over a cache side channel by a normally faulting instruction when it is waiting to commit (delayed fault). %The tool initially sets up each microarchitectural buffer with known values (step 1), executes a test derived by mutating existing attack that contain a faulting load with a consumer that encodes it into a stateful element like the data cache (step 2), extracts the information through Flush+Reload~\cite{Yarom'14} (step 3). Step 3 also includes cross-checking with performance counters. 
Speechminer uses a multi-phased step 2 to identify transient execution violations, wherein step 2(a) causes a race between data access and fault handling while step 2(b) performs the encoding and mispeculation handling. Osiris identifies timing side-channels by running the test on the target twice, once with the trigger and once without. The tests are then clustered based on observed timing difference.

\textbf{Discussion: }Observation-independent fuzzing approaches uncover unintended behaviors without requiring a GRM. However, \textit{the main drawbacks are the need for a large test corpus and the lack of precision in bug localization and classification}. Without a baseline to compare against, distinguishing between expected and erroneous behaviors becomes difficult, particularly when the margin of separation is subtle.

\ignore{
\begin{tcolorbox}[insight]
    \textbf{Limitation:} Not using a GRM or runtime feedback to generate microarchitectural test cases increase the test-corpus, and result in a lack of precision in bug localization and analysis.
\end{tcolorbox}
}
\vspace*{-0.05in}
\subsubsection{Model-Based Relational Testing} \label{sec:typeb-methodology-fuzztesting-mrt}

Model-based Relational Testing (MRT) incorporates abstract models to describe expected relationships between instructions, microarchitectural events, or execution outcomes. In addition to checking against a single correct output, these approaches sometimes assert that certain constraints should hold regardless of the exact output. For example, a model may specify that instruction reordering should not lead to observable timing variations under certain conditions, or that cache behavior must obey specific consistency guarantees. The model used in these approaches could be an ISA emulator enhanced with goal-specific modifications like speculation contracts in Revizor~\cite{revizor-oleksenko} or a projection of a specific microarchitectural component's observation model, like cache observation models in ScamV~\cite{scamv} and ScamV2~\cite{observationrefinement}. %MRT allows for capturing bugs that are relational in nature, such as inconsistencies across program variations or violations of microarchitectural contracts.

Speculation contracts used by Revizor define clauses specifying what an external observer can access (observation clause) and when speculation was allowed (execution clause). Revizor generates a ``contract trace" through a cache side-channel and compares it against the ``contract trace" generated from an instrumented x86 emulator to identify violations of its speculation contracts. When a violation is found, a post-processor creates a minimized version of the violating test case with the sources fenced. %In ScamV, a randomly generated program is run against two inputs that are observationally equivalent according to the model under the test.  %. The model under test is generated by the authors, and its state space reflects the possible values of a program variable at a higher level of abstraction. 
%The indistinguishability of the two inputs is then checked by executing the program and analyzing the side channel. 

ScamV2 \cite{scamv} introduces observation refinement to improve test input generation through a model that is stricter than the baseline model, and in particular captures events that could occur on a side channel of interest. The test program is instrumented with observations for both models. Observation equivalence relations are generated for both the models, and the tests are generated such that the execution with two different inputs is observationally equivalent for the baseline model but not for the refined model, thereby detecting an illegal side channel disclosure. The refined model guides instruction generation by skipping test cases that are not relevant to the model under evaluation and thereby increasing the potential for finding counterexamples.

% REduced the detail shared about revizor and scamv to make the section shorter.
% Revizor~\cite{revizor-oleksenko} uses speculation contracts that define observation and execution clauses. An observation clause would describe the actual values that an external observer should be able to access. For example, \textit{ARCH} (architectural observer) allows for observing the address and data of all loads, the address of stores, and the PCs, thereby enabling the analysis of same address space attacks on caches. An execution clause describes when these observations can be collected. For example, \textit{COND-BPAS} allows observations during speculative execution (CONDitional branch misprediction) and during the speculative execution following a store ByPASs. Revizor additionally generates a contract trace -- a trace of valid observations as permitted by the speculation contract. To evaluate the contract trace collected from the hardware through the cache side channel, Revizor uses an x86 emulator instrumented to generate the contract trace according to the speculation contract. Whenever a violation of the contract is detected by a discrepancy between the hardware trace and the contract trace, the post-processor creates a minimized version of the test with fences around the leakage sources.
\ignore{\begin{tcolorbox}[insight]
    \textbf{Insight~\insightcount:} MRT allows for defining contractual relations between microarchitectural events and execution outcomes.
\end{tcolorbox}}

\textbf{Discussion: }  \textit{The advantage of model relational testing is that it enables the checking of rules without requiring a full hardware model.} However, model construction is non-trivial and may either over-approximate (leading to false positives) or under-approximate (missing real bugs), and it typically requires manual effort or symbolic abstraction of the microarchitecture. For example, Revizor~\cite{revizor-oleksenko}, ScamV~\cite{scamv}, and ScamV2~\cite{observationrefinement} consider the cache to be the side-channel and build models that hinge on that assumption. To port them for detecting violations concerning other side channels, such as TLB or prefetcher tables, a new model would have to be manually created.

\begin{tcolorbox}[insight]
    \textbf{Insight~\insightcount:} Abstract functional/observational-reference models of processor sub-modules enable tractable testing of complex black-box CPUs, despite potential over- or under-approximations of actual behavior.
    % Reliance on an abstract functional/observational-reference model of a sub-module that is easier to build, verify, and use than a complete model of a processor enables easier testing of complex black-box processors even though they could suffer from over-/under-approximations.
    % the idea here is to stress that even though smaller models have their shortcomings, using them over a fully functional CPU model is easier since they can be built and verified quickly, when compared to a functional/observation model of the entire processor. For example, microbenchamarking a cache to build a observation model is much easier and needs significantly less effort when comparing the time needed to build a full CPU model that is "correct". Likewise, a model of a stateful unit can be easily built since observing states is easier than observing a stateless unit.
    
\end{tcolorbox}

\subsubsection{Directed Fuzzing} \label{sec:typeb-methodology-fuzztesting-directed}

In directed fuzzing, the test program mutation is guided towards a specific goal. In microarchitectural fuzzing, the goal is typically a bug in a specific microarchitectural component. ABSynthe~\cite{absynthe-gras} automatically synthesizes contention side-channels in processors. It starts by using a target program and a microarchitecture specification, instruments the target program to synchronize with a spy program created with instructions from the target CPU's leakage map\footnote{Leakage map of an ISA specifies if two instructions or a group of instructions create software visible contention of certain microarchitectural components. This leakage map can also be generated by exhaustively searching the target ISA space.} upon executing a secret dependent operation, uses the raw contention-based measurements from the spy program to repeatedly refine the spy program until the synthesized side-channel can reliably detect the secret. The refinement step uses a Gaussian Na\"ive Bayes classifier to classify the effectiveness of the contention channels. Unlike Covert Shotgun~\cite{covert-shotgun} wherein the possible mutations are limited to a handful of cherry-picked instructions, ABSynthe uses all possible instruction combinations by using a leakage map. ABSynthe is a directed fuzzer since it tries to identify contention side-channels, uses runtime information from the target hardware to refine its spy program, and does not rely on a GRM for runtime information or bug detection. Even though ABSynthe generates a software application-specific spy code, it also identifies possible locations of contention leakage in the microarchitecture.

\ignore{
\begin{tcolorbox}[insight]
    \textbf{Insight:} Directed Fuzzing with adequate feedback is efficient in uncovering bugs but it suffers from its reliance on high-fidelity feedback from the target which is difficult for black-box processors. 
\end{tcolorbox}
}
%% From ABSynthe
%% In the analysis phase, ABSynthe takes a given microarchitecture and the target software as input. It then automatically generates an instrumented binary that synchronizes with a spy program whenever it performs a secret operation. The spy code is initially based on instructions from the target microachitecture’s leakage map. For every well-performing instruction in the leakage map, ABSynthe communicates the raw contention-based measurements to the synthesis engine. The synthesis engine aims to improve the quality of the signal by generating new sequences of instructions based on the contention-based measurements. These new instruction sequences repeatedly refine the spy code until the synthesized side channel can detect the secret information with sufficient confidence. We find in our evaluation that in many cases single instructions can achieve an acceptable performance for side-channel synthesis. In other cases, however, refining the instruction sequence significantly improves the results.

\vspace*{-0.05in}
\subsection{Coverage/Feedback} \label{sec:typeb-dut-feedback-disc}
Similar to ISA fuzzing, black-box CPUs typically offer fewer opportunities for feedback. Revizor uses a prior knowledge-based non-DUT derived feedback metric (pattern coverage) to enhance fuzzing efficiency. Amulet~\cite{amulet-gururaj} overcomes this by implementing Revizor on a architectural simulator, thereby identifying microarchitectural bugs during the design phase on simulators, where the observability is high.

\vspace*{-0.05in}
\subsection{GRMs and Manual Analysis} \label{sec:typeb-grms}
Black-box microarchitectural fuzzing approaches typically don't rely on GRMs~\cite{medusa-moghimi, speechminer-xiao, osiris-weber, scamv, observationrefinement}, primarily due to a dearth of microarchitectural functional models.  %Revizor~\cite{revizor-oleksenko} implements an executable/observation model of the execution contract on an ISA simulator for comparison to the target.  
Most approaches, however, require some manual effort in either the test generation or in analysis. Speechminer~\cite{speechminer-xiao} requires manual creation of test inputs based on the exception lists from manufacturers. ScamV~\cite{scamv} and ScamV2~\cite{observationrefinement} require the manual creation of observation models that are used to automatically generate test inputs. Revizor would also require manual effort in redefining the security policies when the targeted side channel is not the data cache. 

\begin{tcolorbox}[insight]
    \textbf{Insight~\insightcount:} The specificity and complexity of the bugs targeted by \uarch fuzzers require each tool to leverage custom observation models and scalable automated analysis capabilities to be effective.
\end{tcolorbox}
%Finally, RISCVuzz would require manual effort in identifying the root cause of disagreement between multiple black-box CPUs.

%\subsubsection{Differential Fuzzing} \label{sec:typeb-methodology-fuzztesting-diff-fuzzing}

%Differential Fuzzing Approaches compare the behavior of two or more implementations of the same ISA. Differential fuzzing has been discussed extensively in the section detailing ISA-level fuzzing. 

\subsection{Bugs Detected} \label{sec:typeb-bugs}
Microarchitectural fuzzers target a diverse range of vulnerabilities that stem from speculative execution, resource contention, or undocumented hardware behavior.  Medusa~\cite{medusa-moghimi}, SISCloak~\cite{observationrefinement}, Ghostwrite~\cite{riscvuzz} and timing channels by Osiris~\cite{osiris-weber} are some of the new attacks or variants of existing attacks identified by these approaches. Medusa is a Meltdown-style attack that induces unauthorized memory accesses under transient conditions. Transysnther~\cite{medusa-moghimi} also identifies new variants of microarchitectural data sampling (MDS) attacks targeting store/load buffers. SISCloak, detected by ScamV2, exploits speculative loads in an in-order processor.% -- ARM Cortex A53.

%\textcolor{red}{Having an insight box at the end of each fuzzing will be helpful. }
%% type C - RTL fuzzing
\vspace*{-0.05in}
\section{RTL-Level Fuzzing} \label{sec:typec}

%\subsection{Overview} 

Register-Transfer Level (RTL) models implement the functional specification of CPU/IP designs, forming the foundation for silicon tape-out. Fuzzing at the RTL level (pre-silicon) is highly effective due to complete observability and controllability in simulation, permitting fine-grained extraction of coverage metrics and enabling detection of functional bugs. Post-tape-out (post-silicon), hardware designs become a "black box" (as with ISA/\uarch Fuzzing) with limited observability, making debugging challenging. Vulnerabilities discovered post-tapeout through ISA fuzzing (e.g., undocumented instructions) or microarchitectural fuzzing (e.g., timing channels) can be detected by RTL fuzzing~\cite{borkar2024whisperfuzz, sigfuzz} if the source RTL were available---which is not the case for proprietary x86 and ARM processors that ISA and microarchitectural fuzzers predominantly target.

% For example, speculative execution cannot be flagged as a bug when it is the intended behavior, whereas timing variations in constant-time cryptographic designs constitute specification violations and can be detected. The effectiveness of RTL fuzzing is thus fundamentally bounded by the completeness and precision of the functional specification.

% Register-Transfer Level (RTL) models implement the functional specification of CPU/IP designs, forming the foundation for silicon tape-out. Fuzzing at the RTL level (pre-silicon) is highly effective due to complete observability and controllability in simulation, permitting fine-grained extraction of coverage metrics, enhancing verification thoroughness, and enabling the detection of functional bugs. RTL fuzzing can identify design deviations from specification that can manifest as undocumented instructions and timing-channels that ISA fuzzing and \uarch fuzzing try to find post-tapeout. 
% Unlike RTL, which can be easily patched, silicon patches or respins are costly and time-consuming.
The core idea in RTL fuzzing is to automatically generate input stimuli, typically instruction streams or signal patterns, that aim to maximize design coverage. Test coverage is improved by selectively mutating the input stimuli with mutation operators (Appendix~\ref{sec:afl}). RTL fuzzers often require a GRM to identify bugs in a pre-silicon DUT.  %GRMs can be either physical hardware or software-based simulators, each posing trade-offs.%Based on how the fuzzing methodology is carried out, we classify RTL fuzzers into three subclasses as shown in Figure \ref{fig:classification}.

\vspace*{-0.05in}
\subsection{Fuzzing Targets and Input Stimulus} \label{sec:typec-target-gbfuzz}
RTL fuzzers aim to identify bugs and vulnerabilities across a diverse range of designs, including CPU cores~\cite{Kande'22-thehuzz,Canakci'23-processorfuzz} and various IP peripherals~\cite{Laeufer'18}. A common characteristic among all RTL fuzzers is that the source code is available, enabling direct access to design internals during verification. While many RTL fuzzers have been extensively evaluated on designs such as the CVA6~\cite{ariane} and BOOM~\cite{boom} cores, they remain relatively small compared to commodity processors, which often feature out-of-order superscalar execution and large instruction windows~\cite{chipsandcheeseLionCove}. 

% \textbf{Since verification and fuzzing are timing bound, the scalability of RTL fuzzing with increase in design complexity remains unaddressed.}

%RTL fuzzing adopts coverage-driven grey-box fuzzing~\cite{Bohme'17}, which requires the availability of the RTL source code in order to use simulation-time coverage metrics like branch, MUX-toggle, register, and FSM coverage for feedback. Thus, academic works surveyed in this paper are limited to using small open-source processors as targets for their techniques. Although cores like CVA6~\cite{ariane} and BOOM~\cite{boom} are proven on silicon~\cite{boom-silicon, ariane}, they are small compared to commodity general-purpose, server, and even embedded cores that are known to feature out-of-order superscalar processing with large instruction windows \cite{chipsandcheeseLionCove}. Since verification and fuzzing are timing bound, the scalability of RTL fuzzing with increase in design complexity remains unaddressed. 

%\subsection{Input Stimulus} \label{sec:typec-input-gen}

\begin{table*}[!htbp]
\centering
\resizebox{\linewidth}{!}{%

\begin{tabular}{|l|c|lllllll|}
\hline
\multicolumn{1}{|c|}{\multirow{3}{*}{Vulnerabilities in CVA6~\cite{ariane}}}   & \multicolumn{1}{l|}{\multirow{3}{*}{CWE \#}} & \multicolumn{7}{c|}{\# Test Cases\textsuperscript{\Cross}}                                                                                                                                                                                                                                          \\ \cline{3-9} 
\multicolumn{1}{|c|}{}                                   & \multicolumn{1}{l|}{}                        & \multicolumn{1}{c|}{\multirow{2}{*}{TheHuzz\textsuperscript{S}~\cite{Kande'22-thehuzz}}} & \multicolumn{3}{c|}{TheHuzz\textsuperscript{S} Baselines in ~\cite{hybrid-hypfuzz, psofuzz, mabfuzz}}                                                                                                           & \multicolumn{1}{c|}{\multirow{2}{*}{HypFuzz\textsuperscript{H}~\cite{hybrid-hypfuzz}}} & \multicolumn{1}{c|}{\multirow{2}{*}{PSOFuzz\textsuperscript{D}~\cite{psofuzz}}} & \multicolumn{1}{c|}{\multirow{2}{*}{MABFuzz\textsuperscript{D}~\cite{mabfuzz}}} \\ \cline{4-6}
\multicolumn{1}{|c|}{}                                   & \multicolumn{1}{l|}{}                        & \multicolumn{1}{c|}{}         & \multicolumn{1}{c|}{HypFuzz~\cite{hybrid-hypfuzz}}       & \multicolumn{1}{c|}{PSOFuzz\cite{psofuzz}}         & \multicolumn{1}{c|}{MABFuzz\cite{mabfuzz}}         & \multicolumn{1}{c|}{}                         & \multicolumn{1}{c|}{}                     & \multicolumn{1}{c|}{}                     \\ \hline
V1: Incorrect decoding of FENCE.I instruction            & 440                                          & \multicolumn{1}{l|}{$1.36\times10^{4}$} & \multicolumn{1}{l|}{$1.36\times10^{4}$} & \multicolumn{1}{l|}{$6.80\times10^{1}$} & \multicolumn{1}{l|}{$6.00\times10^{2}$} & \multicolumn{1}{l|}{$1.08\times10^{4}$} & \multicolumn{1}{l|}{$1.80\times10^{1}$} & $4.6\times10^{1}$ \\ \hline
V2: Failure to detect cache coherency violation          & 1202                                         & \multicolumn{1}{l|}{$1.72\times10^{5}$} & \multicolumn{1}{l|}{$1.72\times10^{5}$} & \multicolumn{1}{l|}{$1.83\times10^{2}$} & \multicolumn{1}{l|}{$1.20\times10^{3}$} & \multicolumn{1}{l|}{$2.15\times10^{5}$} & \multicolumn{1}{l|}{$1.20\times10^{1}$} & $4.93\times10^{2}$ \\ \hline
V3: Some illegal instruction can be excecuted            & 1242                                         & \multicolumn{1}{l|}{$1.81\times10^{6}$} & \multicolumn{1}{l|}{$1.81\times10^{6}$} & \multicolumn{1}{l|}{$9.30\times10^{1}$} & \multicolumn{1}{l|}{$1.48\times10^{3}$} & \multicolumn{1}{l|}{$1.43\times10^{5}$} & \multicolumn{1}{l|}{$7.40\times10^{1}$} & $1.77\times10^{2}$ \\ \hline
V4: Incorrect exception type in instruction queue        & 1202                                         & \multicolumn{1}{l|}{$4.02\times10^{4}$} & \multicolumn{1}{l|}{$4.02\times10^{4}$} & \multicolumn{1}{l|}{$5.55\times10^{3}$} & \multicolumn{1}{l|}{$2.39\times10^{2}$} & \multicolumn{1}{l|}{$7.40\times10^{3}$} & \multicolumn{1}{l|}{$5.39\times10^{2}$} & $4.0\times10^{1}$ \\ \hline
V5: Incorrect exception while accessing invalid addrress & 1252                                         & \multicolumn{1}{l|}{N/A}        & \multicolumn{1}{l|}{N/A}      & \multicolumn{1}{l|}{$3.52\times10^{2}$} & \multicolumn{1}{l|}{$2.50\times10^{0}$} & \multicolumn{1}{l|}{$2.67\times10^{1}$} & \multicolumn{1}{l|}{$1.38\times10^{2}$} & $3.96\times10^{0}$ \\ \hline
V6: Incorrect decoding of multipy instructions           & 440                                          & \multicolumn{1}{l|}{N/A}        & \multicolumn{1}{l|}{N/A}      & \multicolumn{1}{l|}{$2.71\times10^{3}$} & \multicolumn{1}{l|}{N/A}         & \multicolumn{1}{l|}{$9.80\times10^{4}$} & \multicolumn{1}{l|}{$4.51\times10^{2}$} & N/A \\ \hline
V7: Accessing unimplemented CSRs returns X-values        & 1281                                         & \multicolumn{1}{l|}{N/A}        & \multicolumn{1}{l|}{N/A}      & \multicolumn{1}{l|}{$4.75\times10^{3}$} & \multicolumn{1}{l|}{$1.41\times10^{2}$} & \multicolumn{1}{l|}{$2.08\times10^{3}$} & \multicolumn{1}{l|}{$8.56\times10^{2}$} & $5.97\times10^{1}$ \\ \hline
\end{tabular}
}
\scriptsize
\textsuperscript{\Cross} We use number of test cases as it is hardware platform independent given the same starting seed and fuzzing algorithm.
\newline \textsuperscript{S}: Static Input Scheduling, \textsuperscript{H}: Hybrid Input Scheduling, \textsuperscript{D}: Dynamic Input Scheduling \raggedright

% The versions of TheHuzz shown here are results from different papers and not improvements of the original version. v1: TheHuzz, v2:HypFuzz, v3: PSOFuzz, v4:MABFuzz. 

\caption{Number of test cases generated by fuzzers to detect a subset of CWE Vulnerabilities in CVA6\cite{ariane}. The variance in test cases generated by TheHuzz baselines used in ~\cite{hybrid-hypfuzz, psofuzz, mabfuzz} is shown along with the orginial TheHuzz results. All data is collected from the respective papers.}
\label{tab:typec-input-eval}
\vspace*{-0.1in}
\end{table*}

As shown in Table~\ref{fig:classification},  RTL fuzzers use different types of input stimulus depending on the target at hand.  Since a significant chunk of the work systematized in this paper~\cite{Hur'21-difuzzrtl, Kande'22-thehuzz, hybrid-hypfuzz, Canakci'23-processorfuzz, Hossain'23-socfuzzer, Farzana'19-formalfuzzer, borkar2024whisperfuzz, cascade-razavi, feedbackfuzz, sigfuzz} fuzz CPU designs, the straightforward input stimulus type is assembly instruction sequences (R,A, and X in the table). On the contrary, series of bits~(B)~\cite{Laeufer'18, Muduli'20-hyperfuzzing, Trippel'22-hw-as-sw} serve as a lower-level stimuli to RTL IPs since they do not have an instruction interface like CPUs.

\textbf{Na\"ive and Static Seed Generation: } The naïve approach is to randomly generate and mutate test inputs based on the observed coverage variation. TheHuzz~\cite{Kande'22-thehuzz}, DiFuzzRTL~\cite{Hur'21-difuzzrtl}, ProcessorFuzz~\cite{Canakci'23-processorfuzz} employ this mechanism. \textit{Since the number of potential instructions exposed by the ISA is extremely large, achieving sufficient coverage of the design space to generate effective input seeds becomes a tedious and resource-intensive task.} Typically mutation operators are statically scheduled in a fixed first-in–first-out (FIFO) order, and input seeds are chosen from the database in the same static order, without dynamic priorities.
Random generation also suffers from its inability to effectively identify deeply buried bugs without sufficient guidance as discussed in ISA Fuzzing~(Section~\ref{sec:typea-target-dut-feedback}) and \uarch Fuzzing~(Section~\ref{sec:typeb-dut-feedback-analysis}). Therefore, Cascade~\cite{cascade-razavi} and FeedbackFuzz~\cite{feedbackfuzz} use structured random generation and validate the generated inputs against a ISA simulator. Cascade also uses this step to extract register values and FeedbackFuzz imposes extra structure by incentivizing tests that trigger speculative execution.

\textbf{Dynamic Heuristic-Based Approaches:} Dynamic approaches prioritize high-yield mutation operators (PSOFuzz~\cite{psofuzz}) or  select seeds based on coverage variation during the campaign~\cite{mabfuzz}. PSOFuzz accelerates fuzzing by prioritizing high-yield mutation operators and achieves faster coverage compared to static mutation scheduling. MABFuzz dynamically updates seed selection based on coverage improvement using a Multi-Armed Bandit based approach. MABFuzz balances input space exploration and exploitation (selecting historically well-performing seeds). Adaptive balance in seed-selection improves efficiency by leveraging known information while still exploring novel inputs.

% PSOFuzz \cite{} employs \textbf{particle swarm optimization (PSO), based seed generation to dynamically prioritize mutation operators}. 

% \textbf{ISA Presimulation–Guided Generation:}Beyond PSO and MAB approaches, Cascade~\cite{cascade-razavi} and FeedbackFuzz~\cite{feedbackfuzz} use ISA presimulation to guide input generation. Cascade executes basic blocks on an ISA simulator to extract register values, while FeedbackFuzz extends this by introducing disorder to encourage speculative execution. However, both approaches inherit ISA-level limitations: the large instruction space makes exhaustive search expensive, and filtering for non-faulting instructions, as in Cascade, reduces coverage and may miss corner-case bugs. 

\textbf{Assertion/Property-based Seed Generation:} HypFuzz~\cite{hybrid-hypfuzz} employs hybrid seed generation by combining na\"ive input generation with developer-provided System Verilog Assertions (SVAs) that guide test creation towards uncovered/critical regions of the design~\cite{hybrid-hypfuzz, Farzana'19-formalfuzzer}. HyperFuzzing~\cite{Muduli'20-hyperfuzzing} uses HyperPLTL properties with bit inputs to capture high-level security relations across multiple traces. \textit{Reliance on assertions and hyperproperties could be problematic for fuzzer scaling, as there is no automated way of generating them.} Manual insertion of assertions can also suffer from human bias leading to missing corner-case bugs. 

% \textbf{Moreover, the number of required assertions grows exponentially with design size, making such approaches difficult to scale for large and complex hardware systems.}

\textbf{ML-based Seed Generation:}
Recent research in ML-based seed generation has gained significant traction, with approaches leveraging machine learning to produce smarter inputs that improve coverage and bug detection. ChatFuzz~\cite{chatfuzz} leverages a Large Language Model~(LLM)~\cite{gpt-radford}, trained on disassembled binaries and refined through Reinforcement Learning~(RL)~\cite{rl-introduction-sutton} to enforce syntactic and semantic validity, for generating test inputs. During fuzzing, the reinforcement learning model incorporates coverage feedback to guide the LLM. HFL~(Hardware Fuzzing Loop with RL)~\cite{hfl,rlfuzz} employs Long Short-Term Memory~(LSTM)~\cite{lstm-schmidhuber} and RL for input generation, to overcome RTL simulation bottlenecks. GenHuzz~\cite{genhuzz} combines LLMs with RL to generate instruction sequences that better explore the large CPU state space.

\textbf{Discussion:} %As illustrated in Table~\ref{tab:typec-input-eval}, we systematize the efficacy of different seed generation mechanisms on CVA6 and extract the common vulnerabilities addressed across these works. 
In Table~\ref{tab:typec-input-eval}, we show the number of tests TheHuzz, HypFuzz, PSOFuzz, and MABFuzz generate to identify the same seven vulnerabilities on CVA6. \textit{Hybrid and dynamic seed scheduling algorithms are effective at identifying bugs quicker due to their prioritization of interesting seeds}. This is evident by the number of tests generated by HypFuzz, PSOFuzz, and MABFuzz being lower than the original TheHuzz~\cite{Kande'22-thehuzz} on average. 

\begin{tcolorbox}[insight]
\textbf{Insight~\insightcount:} Well crafted input seeds and seed-generation mechanisms directly contribute to the efficiency and efficacy of fuzzers.
% Efficiency and Efficacy of fuzzing increases as we move from static, to hybrid and dynamic seed generation. 
% Static seed generation is the slowest, hybrid approaches outperform static methods, and dynamic strategies are faster than both. 
\end{tcolorbox}

Table~\ref{tab:typec-input-eval} also presents two critical concerns affecting the progress of RTL fuzzing research. First, widely used baselines like TheHuzz are not publicly available for comparative studies. This lack of reproducibility prompts authors to re-implement baselines with publicly available information leading to wide discrepancies in published results (columns 3 vs 4,5,6 in Table~\ref{tab:typec-input-eval}). Second, there is a lack of standardized metrics that get reported for all fuzzers. For instance, some fuzzers report runtime speedup and time to bug over platform invariant metrics like the number of test cases generated to demonstrate effectiveness. This consequently hampers the ability to comparatively interpret the effectiveness of a fuzzers reporting different metrics. We discuss these issues further in Section~\ref{sec:typec-coverage-disc}.

\vspace*{-0.05in}
\subsection{Fuzzing Methodology} \label{sec:typec-input-methodology}
We classify RTL fuzzing into two classes based on their fuzzing methodology: (a) Fuzzing Hardware as software (FHaS), and (b) Fuzzing Hardware as Hardware (FHaH). 
%\textcolor{red}{Similar to previous sections, have a leading paragraph stating there are 3 subclases in fuzzing.Done!}

% \vspace*{-0.05in}
%\textbf{Adoption of Software Fuzzers for Hardware (SFH).} 

\subsubsection{Fuzzing Hardware as Software (FHaS)}

Fuzzing Hardware as Software (FHaS) translates RTL designs to software models (e.g., C++ via Verilator~\cite{Verilator}) and applies software fuzzing frameworks like AFL~\cite{AFL'23} or libFuzzer~\cite{libfuzz}. RFuzz~\cite{Laeufer'18} pioneered this approach by injecting fuzzed inputs into the translated design's memory and using code coverage and crashes as feedback to detect semantic inconsistencies and logic faults through System Verilog Assertions (SVAs). The major shortcoming of FHaS is that software-like based RTL simulators~\cite{Verilator, iverilog} lack support for RTL innate features like \texttt{x} (undefined) and \texttt{z} (impedance) states and do not support the complete language standard. RFuzz and DiFuzzRTL~\cite{Hur'21-difuzzrtl} while classified as FHaS also support fuzzing target designs on FPGAs, outperforming simulator based fuzzers. Trippel et al.~\cite{Trippel'22}, HyperFuzzing~\cite{Muduli'20-hyperfuzzing},  Cascade~\cite{cascade-razavi}, and FeedbackFuzz~\cite{feedbackfuzz} fall into this category.

\begin{tcolorbox}[insight]
   \textbf{Limitation~\limitationcount:} 
   Adapting software fuzzers to hardware faces fundamental challenges: hardware lacks crash semantics, requires sequential stimuli, and depends on HDL-specific constructs and coverage models incompatible with software fuzzing abstractions.
   %Adapting software fuzzers such as AFL to hardware exposes a fundamental abstraction gap: hardware lacks crash semantics, requires sequential stimuli, and depends on HDL constructs and coverage models that software fuzzers cannot represent.
\end{tcolorbox}

\subsubsection{Fuzzing Hardware as Hardware (FHaH)}

Fuzzing Hardware as Hardware refers to fuzzing the hardware at the HDL level through EDA tools or through FPGA emulations, in contrast to FHaS. Recent advances in hardware fuzzing have embraced this domain-specific approach, fuzzing hardware as hardware, to eliminate RTL translation and the associated preliminary tasks and equivalence checks. In this approach, the design is fuzzed at the hardware's native abstraction level using industry-standard EDA tools, such as Synopsys VCS, Cadence Xcelium, Aldec Riviera or Siemens Questa~\cite{synopsys-vcs, cadence-xcelium, siemens-questa, Aldec}. \textit{EDA tools provide an accurate view of real hardware behavior, including timing-sensitive faults and low-level state transitions.} The inputs are assembly test sequences that are compiled and loaded into design memory before simulation. The inputs are simultaneously tested on a GRM (ISA simulator), and their execution traces are compared to detect bugs and vulnerabilities. As we discuss in Section~\ref{sec:typec-coverage-disc}, the coverage metrics extracted from the RTL simulators can capture intrinsic hardware characteristics. The majority of the RTL fuzzers, such as TheHuzz~\cite{Kande'22-thehuzz}, DiFuzzRTL~\cite{Hur'21-difuzzrtl}, PSOFuzz~\cite{psofuzz}, MABFuzz~\cite{mabfuzz}, and HypFuzz~\cite{hybrid-hypfuzz}, fall under this fuzzing methodology. ProcessorFuzz~\cite{Canakci'23-processorfuzz}, unlike other RTL fuzzers, uses Control Status Register~(CSR) Transitions (Section~\ref{sec:typec-coverage-disc}) as a proxy for coverage to mutate effective seeds. More recently, approaches exploring graph-based~\cite{graphfuzz} and sequence-learning models to guide test generation and coverage prediction~\cite{hfl} have been proposed. Finally, RTL fuzzers that rely on GRMs inherently are limited by the GRM  trustworthiness as discussed in Section~\ref{sec:typea-grms} and Section~\ref{sec:typeb-grms}.

%\textbf{Notably, GraphFuzz~\cite{graphfuzz} leverages graph neural networks to mimic RTL designs in graph form, enabling faster RTL simulations while preserving the intrinsic hardware characteristics and semantic details of the design.} In addition, to address the bottleneck of slow RTL simulations, HFL~\cite{hfl} leverages Long Short-Term Memory (LSTM) networks, widely used in Natural Language Processing (NLP), in tandem with Reinforcement Learning (RL) to design a hardware coverage predictor by learning the design semantics. The ML-based coverage predictor eliminates the need for full RTL simulation, enabling faster feedback and guiding fuzzing more efficiently. These approaches highlights the growing potential of combining traditional EDA-based flows with ML-driven abstractions to address the scalability bottlenecks of RTL fuzzing. \textcolor{red}{Is HFL abbreviated?}

\begin{tcolorbox}[insight]
\textbf{Tradeoff~\tradeoffcount:} While FHaH preserves intrinsic hardware characteristics, it remains constrained by RTL simulation performance.
% impact due to  high runtime-overhead instrumentation by sophisticated simulators.
\end{tcolorbox}

\vspace*{-0.05in}
\subsection{Coverage Metrics} \label{sec:typec-coverage-disc}

RTL fuzzers leverage diverse coverage metrics~\cite{verificationacademyCoverage}, each with trade-offs in scalability, precision, and bug detection (Table~\ref{tab:comparison}, Appendix~\ref{tab:appendix-coverage-metrics}). RFuzz pioneered the field with MUX-based coverage, but it suffers from high instrumentation overheads and fails to scale to larger designs like BOOM or Mor1kx. Moreover, MUX coverage is clock-insensitive (i.e, it does not capture the interplay between toggling events across clock cycles). Therefore, RFuzz fails to identify bugs in FSM state transitions and in multiplexers implemented with basic gates, which DiFuzzRTL~\cite{Hur'21-difuzzrtl} and TheHuzz~\cite{Kande'22-thehuzz} can uncover.

In FHaS, RTL is translated into an executable binary, causing hardware-specific coverage metrics (FSM, branch, toggle) to be lost.  While edge coverage in software is closely analogous to branch or MUX coverage, it is not scalable for large designs. In contrast, FHaH preserves intrinsic RTL behavior, providing relevant coverage metrics.

DiFuzzRTL improves scalability by introducing register coverage, reducing overhead to 7\% compared to 71\% for TheHuzz. 
ProcessorFuzz reduces coverage overhead by introducing CSR-transition coverage, which is agnostic to RTL simulators and requires no instrumentation. However, both approaches are limited to registers driving multiplexer select signals and therefore miss bugs such as illegal debug access. TheHuzz, by contrast, can expose such fine-grained bugs at the cost of substantial instrumentation overheads imposed by EDA tools. The runtime overheads for TheHuzz are 71\% higher than DiFuzzRTL~\cite{Kande'22-thehuzz} due to design instrumentation by the EDA tool. Moreover, RTL simulations are also estimated to be 79$\times$ slower than ISA simulations~\cite{Canakci'23-processorfuzz}.

Formal-verification aided fuzzing approaches such as HypFuzz~\cite{hybrid-hypfuzz} and SoCFuzzer~\cite{Hossain'23-socfuzzer} are effective at reaching hard-to-cover regions, but they do not scale easily to large designs. Moreover, formal tool coverage can have human bias, and the coverage achieved on larger designs is often limited, raising questions about their practicality. More recently, ML-based approaches, such as HFL~\cite{hfl}, leverage LSTM-based coverage predictors to accelerate fuzzing. However, their accuracy remains below that of EDA tools, and they risk producing false positives. Similarly, graph-based models that rely on toggling nodes to mimic design activity face scalability challenges, making them impractical for very large hardware systems. These tradeoffs underscore the need for hybrid and adaptive coverage metrics that balance accuracy, scalability, and overhead.

\textbf{Discussion}: A pervasive limitation across RTL fuzzing literature is the lack of contextual information about reported results, hindering standardization and interpretability. For example, neither TheHuzz nor ChatFuzz provides breakdowns of their coverage points. For instance, in designs like Rocket Core~\cite{Asanovic'16-rocket} with $\sim$600K coverage points, it remains unclear how many FSM, expression, statement, block, branch or toggle coverage points contribute to reported percentages. This opacity extends broadly. While PSOFuzz~\cite{psofuzz} and MABFuzz~\cite{mabfuzz} report coverage improvements, these gains are not contextualized against concrete metrics such as total branches, FSM states, or toggle points, making it impossible to assess whether improvements represent genuinely new design space exploration or measurement artifacts. Similarly, GenHuzz~\cite{genhuzz} reports executing over 10K tests per second and outperforming Cascade and ChatFuzz, yet does not specify the experimental conditions under which this throughput was achieved—unlike Cascade's comprehensive sweep study. ChatFuzz~\cite{chatfuzz} claims 97.02\% condition coverage in 49 minutes compared to TheHuzz's 75\% in 30 hours, but without standardized coverage definitions or experimental parameters, such comparisons remain difficult to validate or reproduce. Furthermore, higher coverage percentages do not necessarily correlate with bug detection efficiency—even with reported improvements of 0.3–2.2\%, absolute coverages remain below 70\%, raising scalability concerns for larger designs. These inconsistencies highlight a critical gap: the absence of community-wide standards for coverage definitions, measurement methodologies, and reporting practices. Given that coverage-driven verification is fundamental to RTL fuzzing—serving as both the primary feedback mechanism and effectiveness metric—establishing standardized benchmarks, metrics, and evaluation protocols is essential for reproducible research and meaningful cross-tool comparisons.

\begin{tcolorbox}[insight]
\textbf{Call to the Community:\\} 
\textbf{Standardize:} Adopt uniform coverage metrics, measurement protocols, and experimental reporting to enable reproducible comparisons. \\
\textbf{Open Source:} Release fuzzing tools, benchmarks, and datasets to accelerate collaborative progress.
\end{tcolorbox}

\subsection{Bug Detection and Analysis} \label{sec:typec-manual-analysis}

RTL fuzzers predominantly identify functional bugs, identified as inconsistencies~(\textit{Inc}) between the target RTL and the GRM~\cite{Hur'21-difuzzrtl, Kande'22-thehuzz, hybrid-hypfuzz, Canakci'23-processorfuzz, borkar2024whisperfuzz, cascade-razavi, feedbackfuzz, sigfuzz}. Some fuzzers identify timing side-channels~\cite{sigfuzz, borkar2024whisperfuzz}. Recent co-design approaches further extend this to firmware, detecting bugs at the hardware–software boundary~\cite{Muduli'20-hyperfuzzing, Trippel'22-hw-as-sw}. Next, RTL fuzzers rely on manual analysis to determine the root cause of failing test cases. This is critical to filter false positives. Cascade~\cite{cascade-razavi}, FeedbackFuzz~\cite{feedbackfuzz}, and SIGFuzz~\cite{sigfuzz} automate this step. Cascade iteratively reduces the test program, FeedbackFuzz expedites Cascades iterative reduction with an LLM, and SIGFuzz uses a hash in traces to aid rapid fault localization.  

\vspace*{-0.05in}

\section{Cross-cutting Insights and Opportunities} \label{sec:cross-cutting-insights}

Our systematization reveals several recurring themes that transcend individual fuzzing approaches and abstraction layers, highlighting fundamental challenges and opportunities in hardware fuzzing.

\textbf{Structured Input Space Exploration is better than Random Search.}
Across all fuzzing classes, structured input generation consistently outperforms unconstrained randomization. In ISA fuzzing, SkipScan's structured byte-level mutations prove more effective than pure random instruction generation (Table~\ref{tab:typea-input-effectiveness-undocumentedinsts}), while Revizor's microarchitectural contracts guide exploration toward timing-channel-inducing sequences more efficiently. Similarly, RTL fuzzers with hybrid and dynamic seed generation prioritize interesting input seeds and use feedback from the target design to refine inputs between runs. This pattern underscores a fundamental principle: hardware's vast input space and design complexity require domain knowledge to constrain the search space. Pure randomization, while preferable for exploring the input space, is inefficient given hardware complexity and intricate correctness conditions.

\textbf{Heterogeneous Metrics and Closed-source Fuzzers limit Quantitative Comparison.}
Absence of standardized evaluation metrics significantly impacts interpretability of results. ISA fuzzers report bugs found or undocumented instructions discovered; microarchitectural fuzzers measure leakage rates or attack variants; RTL fuzzers cite improvement in coverage percentages without specifying coverage types (FSM, toggle, branch), present identified bugs, or present the number of instructions generated to identify a bug. Even within a single class, reported metrics vary. Moreover, the lack of open-source fuzzers leads to non-trivial differences in baseline metrics reported by different publications comparing against the same baseline (Table~\ref{tab:typec-input-eval}). This heterogeneity permits only qualitative comparisons and obscures genuine advances. The community needs standardized benchmarking corpora, unified metric definitions, and transparent reporting protocols.

\textbf{Attack-Driven \uarchn itectural Fuzzing.}
A fundamental divide exists in fuzzing objectives across abstraction layers. ISA fuzzing is \textit{specification-driven}, seeking deviations from documented architectural contracts. RTL fuzzing is \textit{coverage-driven}, maximizing exploration of design state space to expose functional bugs. In contrast, \uarchn itectural fuzzing is predominantly \textit{attack-driven}. Tools are engineered around known side-channel and transient-execution attack templates (e.g., Spectre, Meltdown variants). This limits the reach of \uarchn itectural fuzzers resulting in discoveries of variants within known vulnerability classes rather than uncovering entirely new \uarchn itectural failure modes. Moving beyond template-based approaches toward systematic exploration of the \uarchn itectural state space remains an open challenge.

% Osiris~\cite{osiris} and Transynther~\cite{medusa} demonstrate this pattern: despite running for 500+ and 26 CPU hours respectively, they synthesize variations of existing attacks (new Flush+Reload variants, Medusa as a ZombieLoad subtype) rather than novel vulnerability classes. 

\textbf{Conservation of Manual Effort.}
While hardware fuzzing significantly improves test coverage and bug detection rates compared to manual testing, it does not eliminate manual effort. Rather, it redistributes it. While automated test generation through fuzzing reduces manual test writing, it increases demands on oracle/GRM construction, configuration specification, coverage extraction, and result triage. ISA fuzzers require precise ISA specifications and reference implementations; microarchitectural fuzzers need attack templates and leakage models; RTL fuzzers depend on assertions, coverage configurations, and bug classification logic. Differential fuzzing can tolerate buggy oracles/GRMs by testing multiple targets and comparing results across them. This would require access to multiple processors which is impractical within single-vendor design teams. 

% Quantifying and comparing manual effort across approaches remains difficult due to inconsistent reporting of setup time, configuration complexity, and false positive rates.

\textbf{Limited Algorithmic Diversity.}
Despite the proliferation of hardware fuzzers, underlying search algorithms show surprising homogeneity. Most RTL-fuzzers employ variations of coverage-guided fuzzing (AFL-style), genetic algorithms, or reinforcement learning—often as incremental refinements rather than fundamentally new strategies. Hardware fuzzing can benefit from exploring alternative search paradigms: constraint solving for targeted property violation, program synthesis for oracle generation, or causal inference for root-cause localization. The algorithmic monoculture risks diminishing returns as each refinement yields marginal improvements.

%ISA fuzzers tweak mutation operators; RTL fuzzers adjust feedback signals; microarchitectural fuzzers modify seed scheduling. 

\textbf{Trusted Golden Reference Models.}
Effective fuzzing requires correct behavioral specification or reference implementation. Constructing correct Golden Reference Models (GRMs) remains challenging, particularly for complex microarchitectural behaviors and proprietary designs. ISA fuzzers often rely on differential testing across implementations, but this only detects \textit{inconsistencies}, not violations of intended behavior when all implementations share the same bug. Microarchitectural fuzzers face even greater challenges: no specifications exist for cache replacement policies, speculative execution boundaries, or pipeline interactions. Manual GRM construction (e.g., ScamV~\cite{scamv}) is labor-intensive and error-prone, while automatically derived models from HDL suffer from over/under-approximation. Target heterogeneity, diverse ISA extensions, vendor-specific optimizations, and generational differences further exacerbate GRM construction difficulties. Differential fuzzing across multiple implementations partially mitigates this, but remains impractical for single-vendor design teams as discussed in 'Conservation of Manual Effort' above. Automating accurate GRM generation from specifications or synthesizing approximate-yet-useful behavioral models represents a critical research direction for both fuzzing and conventional verification.

While these limitations present significant challenges, they also represent clear opportunities. In the following section, we outline specific directions that address each of these fundamental barriers.

\section{Future Directions} \label{future-directions}

The cross-cutting insights presented in Section~\ref{sec:cross-cutting-insights} reveal fundamental limitations that currently prevent hardware fuzzing from matching the maturity and effectiveness of software fuzzing. 
In this section, we outline a forward-looking research agenda aimed at addressing these barriers. 
Each direction tackles a specific limitation identified in our systematization while opening pathways for concrete research projects, together forming a roadmap toward more systematic and effective hardware fuzzing.

% Despite the progress made in recent years, hardware fuzzing remains hampered by fundamental limitations as discussed in Section~\ref{sec:cross-cutting-insights} that prevent it from matching the maturity and effectiveness of its software counterparts. 
% Current approaches often rely on random or lightly constrained input generation, incomplete reference models, or narrow single-layer testing methodologies, leaving wide swaths of the hardware design space unexplored. 
% In this section, we outline a forward-looking research agenda aimed at overcoming previously discovered barriers. Each direction addresses a core limitation while also opening pathways for concrete projects, together forming a roadmap toward more systematic and effective hardware fuzzing.

\textbf{Intelligent Input Generation.}
One of the most pressing challenges lies in input generation. 
% Existing fuzzers largely depend on random or heuristically mutated instruction streams, which rarely penetrate the deep microarchitectural states where elusive bugs hide. 
We need intelligent approaches that blend randomness with domain-specific knowledge. Given the proliferation of AI, a promising direction is to leverage code generation capabilities of large language models (LLMs), trained on ISA specifications and test suites, to generate semantically rich instruction sequences that stress edge cases such as speculation boundaries, misaligned memory operations, or rare exception paths. RTL Fuzzers~\cite{chatfuzz, rlfuzz, genhuzz} have successfully attempted to leverage LLMs for RTL fuzzing. This approach can be extended to intelligently generate tests for targets with limited feedback, such as black-box commodity processors. %Future work can leverage LLMs to intelligently generate tests for targets with limited feedback such as the black-box commodity processors.

Beyond just generating code, AI agents could also be deployed to guide the fuzzing process itself. These agents could be trained to observe the fuzzer's progress and reward input sequences that expand state-space coverage. This would help the fuzzer move beyond shallow testing and systematically probe the deep, complex interactions most likely to harbor bugs. This approach turns fuzzing from a noisy, brute-force process into a more targeted, guided exploration.

\textbf{Reliable and Scalable Reference Models.}
Our analysis reveals that even the best-crafted input is useless without a reliable oracle to evaluate correctness.  Traditional fuzzing depends on golden reference models (GRMs), yet building cycle-accurate models for modern out-of-order, multicore processors is impractical.  Scalable alternatives are urgently needed.  One direction is to build layered reference models that combine lightweight architectural simulators with focused micro-models for components like caches, branch predictors, or TLBs, enabling targeted bug localization such as the models developed in Amulet~\cite{amulet-gururaj} that make the first foray in this direction. Another potential direction is to pursue differential fuzzing, comparing outputs across multiple implementations or steppings, avoiding reliance on a single trusted reference. A third promising approach involves the use of learned oracles, where AI models trained on large execution traces predict expected behaviors with probabilistic confidence. Together, these directions shift the bottleneck from monolithic GRMs toward a spectrum of scalable oracles, allowing fuzzing campaigns to operate at real-world scales.

In many cases, especially when proprietary designs are tested, full GRMs are unavailable, and fuzzers are forced to rely on coarse oracles that miss subtle deviations. This creates an urgent need for alternative feedback sources.  One direction is to exploit microarchitectural side-effects such as hardware performance counters, cache miss patterns, or branch misprediction statistics as additional feedback signals.  Another direction is to perform side-channel guided fuzzing, where existing attack detection approaches could be combined with fuzzing, such that unusual microarchitectural behaviors are treated as indicators of latent bugs even when architectural results appear correct.  Differential testing across different CPU steppings using such feedback could also surface latent regressions introduced during design evolution.

\textbf{Cross-Layer (Hybrid) Fuzzing.}
Beyond oracles, many vulnerabilities stem not from a single abstraction layer, but from interactions across levels of the hardware stack. For instance, a benign decoder bug may only manifest when compounded with speculative execution, or a cache protocol flaw may surface only under specific ISA-level instruction mixes. Today’s fuzzers rarely traverse these boundaries, remaining siloed at the ISA, microarchitecture, or RTL level. 

A key research direction is cross-layer hybrid fuzzing, where test stimuli, coverage metrics, and oracles are coordinated across layers. For example, ISA-level fuzzers could generate workloads likely to stress decode logic, passing them downstream to \uarchn-level fuzzers, while anomalies discovered at low levels could feed back to seed higher-level generators. To make cross-layer fuzzing work, we need new metrics that span multiple layers. These metrics would track not only how many instructions have been covered but also how much of the microarchitectural event space has been explored (e.g., how many different types of cache miss patterns have been triggered) and how much of the RTL state has been reached. This provides a common language for these vertical fuzzing pipelines to communicate and guide each other.  By bridging layers, hybrid approaches could expose the subtle, interdependent flaws that evade today’s fuzzers.

\textbf{Hardware Fuzzing Benchmarks and Standardization}
A unified benchmarking ecosystem would substantially accelerate progress in hardware fuzzing. The community could collaboratively develop a hardware fuzzing counterpart to software fuzzing's FuzzBench~\cite{metzman2021FuzzBench}, built around open-source CPU designs such as Ariane~\cite{ariane}, BOOM~\cite{boom}, and mor1kx~\cite{Morlkx}, along with CAD-for-assurance~\cite{nathCADforAssurance} benchmarks where faults are introduced in a controlled manner using automated bug injection frameworks like Encarsia~\cite{encarsia}. This benchmark suite would be accompanied by a common experimentation harness capable of running fuzzers targeting different abstraction levels—ISA, \uarchn, or RTL. Each fuzzer would produce results in a standardized format, including coverage metrics, bug-detection latency, number of tests generated, and resource utilization. Such a benchmark would make experimental results comparable and transparent, thereby incentivizing improved hybrid methodologies and discouraging bespoke evaluations that hinder cross-fuzzer comparison. Over time, this shared infrastructure would play the same role FuzzBench has in software testing by acting as a neutral proving ground that supports the development of scientifically rigorous, reproducible hardware fuzzing research.

\textbf{Human-in-the-Loop Automation.}
Despite advances in smart input generation, scalable oracles, cross-layer pipelines, and richer feedback, human expertise remains central to effective fuzzing. Current workflows often generate massive volumes of redundant test traces, overwhelming engineers and slowing down bug triage. 

An important research direction is the development of human-in-the-loop fuzzing frameworks that leverage automation to reduce noise while empowering experts to guide exploration. For instance, clustering techniques could automatically group failing traces by root cause, while visualization tools could expose unexplored state-space regions, allowing experts to strategically redirect fuzzing efforts. 
Future systems could also propose candidate tests that engineers could then refine into reusable litmus tests. This turns one-off fuzzing results into enduring validation artifacts, building a library of targeted tests that can be used to prevent regressions in future designs. This approach closes the loop between automated exploration and human insight, making fuzzing both more efficient and more impactful in practice.

\vspace*{-0.05in}
\section{Conclusion}
This paper presents a comprehensive study of contemporary hardware fuzzing approaches applied across multiple layers of abstraction. We systematize the community’s efforts to build efficient hardware fuzzers by introducing a classification based on abstraction levels, alongside a set of design dimensions that characterize each approach. %Specifically, we categorize existing works into ISA-level, microarchitectural-level, and RTL-level fuzzers, discuss the trade-offs inherent to each, and critically analyze them across the proposed design dimensions. 
Our analysis yields key insights into the current state-of-the-art and highlights open research challenges. We also provide a flowchart~(Figure~\ref{fig:appendix-flowchart}) that can be used in conjunction with Table~\ref{tab:comparison} to choose between multiple fuzzers. %We conclude by proposing future directions aimed at advancing the development of more effective and efficient hardware fuzzers. This work is intended to serve as a foundational reference for researchers and practitioners aiming to build the next generation of hardware fuzzers.

We find that contemporary hardware fuzzing approaches have evolved from brute force random processes into a more sophisticated, albeit still limited, discipline, incorporating targeted approaches but remaining constrained by incomplete reference models and siloed testing methodologies. We believe that by enabling intelligent instruction stream generation, developing scalable and layered hardware oracles, standardized evaluation metrics and benchmarks, and enabling cross-layer testing, we can enable systematic and comprehensive hardware verification.%, ultimately bridging the gap with its software counterparts.

\newpage 
\section*{Ethical Considerations}
Since the contribution of this work is to systematize the state-of-the-art in Hardware Fuzzing, we do not foresee any potential harm to stakeholders from publishing it. We sourced all data for this work from publicly available research publications while complying with their licenses when available.

\section*{Open Science Contribution}
In this work, we survey academic works that have been accepted to conferences through peer-review or publicly archived. Our references section cite all these works. The quantitative data that we present for our analysis is directly derived from the papers and we do not run any experiments of our own. 

{\footnotesize \bibliographystyle{unsrtnat}
\bibliography{ref,references}}

@inproceedings{hybrid-hypfuzz,
author = {Chen, Chen and Kande, Rahul and Nguyen, Nathan and Andersen, Flemming and Tyagi, Aakash and Sadeghi, Ahmad-Reza and Rajendran, Jeyavijayan},
title = {HyPFuzz: formal-assisted processor fuzzing},
year = {2023},
isbn = {978-1-939133-37-3},
publisher = {USENIX Association},
address = {USA},
booktitle = {Proceedings of the 32nd USENIX Conference on Security Symposium},
articleno = {77},
numpages = {18},
location = {Anaheim, CA, USA},
series = {SEC '23}
}

@INPROCEEDINGS{hfl,
  author={Wu, Lichao and Rostami, Mohamadreza and Li, Huimin and Sadeghi, Ahmad-Reza},
  booktitle={2025 Design, Automation \& Test in Europe Conference (DATE)}, 
  title={HFL: Hardware Fuzzing Loop with Reinforcement Learning}, 
  year={2025},
  volume={},
  number={},
  pages={1-7}
 }

@INPROCEEDINGS{kevin,
  author={Gubbi, Kevin Immanuel and Tarighat, Mohammadnavid and Sudarshan, Arvind and Kaur, Inderpreet and Kota, Pavan Dheeraj and Sasan, Avesta and Homayoun, Houman},
  booktitle={2025 26th International Symposium on Quality Electronic Design (ISQED)}, 
  title={State of Hardware Fuzzing: Current Methods and the Potential of Machine Learning and Large Language Models}, 
  year={2025},
  volume={},
  number={},
  pages={1-7},
  doi={10.1109/ISQED65160.2025.11014308}}

@misc{synfuzz,
      title={SynFuzz: Leveraging Fuzzing of Netlist to Detect Synthesis Bugs}, 
      author={Raghul Saravanan and Sudipta Paria and Aritra Dasgupta and Venkat Nitin Patnala and Swarup Bhunia and Sai Manoj P D},
      year={2025},
      eprint={2504.18812},
      archivePrefix={arXiv},
      primaryClass={cs.CR},
      url={https://arxiv.org/abs/2504.18812}, 
}

@INPROCEEDINGS{rlfuzz,
  author={Götz, Raphael and Sendner, Christoph and Ruck, Nico and Rostami, Mohamadreza and Dmitrienko, Alexandra and Sadeghi, Ahmad-Reza},
  booktitle={2025 IEEE International Symposium on Hardware Oriented Security and Trust (HOST)}, 
  title={RLFuzz: Accelerating Hardware Fuzzing with Deep Reinforcement Learning}, 
  year={2025},
  volume={},
  number={},
  pages={358-369},
  }

@misc{graphfuzz,
      title={Accelerating Hardware Verification with Graph Models}, 
      author={Raghul Saravanan and Sreenitha Kasarapu and Sai Manoj Pudukotai Dinakarrao},
      year={2025},
      eprint={2412.13374},
      archivePrefix={arXiv},
      primaryClass={cs.CR},
      url={https://arxiv.org/abs/2412.13374}, 
}

@article{mabfuzz,
title={{MABFuzz: Multi-Armed Bandit Algorithms for Fuzzing Processors}},
author={Gohil, Vasudev and Kande, Rahul and Chen, Chen and Sadeghi, Ahmad-Reza and Rajendran, Jeyavijayan},
journal={Design, Automation and Test in Europe Conference},
year={2024}
}

@INPROCEEDINGS{chatfuzz,
  author={Rostami, Mohamadreza and Chilese, Marco and Zeitouni, Shaza and Kande, Rahul and Rajendran, Jeyavijayan and Sadeghi, Ahmad-Reza},
  booktitle={2024 Design, Automation \& Test in Europe Conference \& Exhibition (DATE)}, 
  title={Beyond Random Inputs: A Novel ML-Based Hardware Fuzzing}, 
  year={2024},
}

@INPROCEEDINGS{psofuzz,
  author={Chen, Chen and Gohil, Vasudev and Kande, Rahul and Sadeghi, Ahmad-Reza and Rajendran, Jeyavijayan},
  booktitle={2023 IEEE/ACM International Conference on Computer Aided Design (ICCAD)}, 
  title={PSOFuzz: Fuzzing Processors with Particle Swarm Optimization}, 
  year={2023},
  volume={},
  number={},
  pages={1-9}
 }

@article{borkar2024whisperfuzz, title={{WhisperFuzz: White-Box Fuzzing for Detecting and Locating Timing Vulnerabilities in Processors}}, author={Borkar, Pallavi and Chen, Chen and Rostami, Mohamadreza and Singh, Nikhilesh and Kande, Rahul and Sadeghi, Ahmad-Reza and Rebeiro, Chester and Rajendran, Jeyavijayan}, journal={arXiv preprint arXiv:2402.03704}, year={2024} }

@INPROCEEDINGS{sigfuzz,
  author={Rajapaksha, Chathura and Delshadtehrani, Leila and Egele, Manuel and Joshi, Ajay},
  booktitle={2023 Design, Automation \& Test in Europe Conference \& Exhibition (DATE)}, 
  title={SIGFuzz: A Framework for Discovering Microarchitectural Timing Side Channels}, 
  year={2023},
  volume={},
  number={},
  pages={1-6},
  doi={10.23919/DATE56975.2023.10136966}}

@misc{ossfuzz,
	title={OSS-Fuzz},
	url={https://google.github.io/oss-fuzz/},
	author  = {Google},
	note = {Last Accessed : 6/1/2024},
}

@misc{libfuzz,
	author = {LLVM 21.0.0 Documentation},
	title = {libFuzzer a library for coverage-guided fuzz testing. 2014; LLVM 21.0.0 - git documentation --- llvm.org},
	howpublished = {\url{https://llvm.org/docs/LibFuzzer.html}},
	year = {},
	note = {[Accessed 11-04-2025]},
}

@misc{ariane,
	Author={Openhwgroup},
	url={https://github.com/openhwgroup/cva6},
	note={Last Accessed : 6/1/2024},
}

@article{Tricheck,
	title={TriCheck: Memory Model Verification at the Trisection of Software, Hardware, and ISA},
	author={Caroline Trippel and Yatin A. Manerkar and Daniel Lustig and Michael Pellauer and Margaret Martonosi},
	journal={Proceedings of the Twenty-Second International Conference on Architectural Support for Programming Languages and Operating Systems},
	year={2016},
	
}

@misc{Aldec,
	author = {Aldec},
	title = {{Riviera-PRO: Advanced Verification Platform}},
	note = {Last Accessed : 6/1/2024},
	url = {https://www.aldec.com/en/products/functional_verification/riviera-pro},
}

@article{genhuzz,
  title={GenHuzz: An Efficient Generative Hardware Fuzzer},
  author={Wu, Lichao and Rostami, Mohamadreza and Li, Huimin and Rajendran, Jeyavijayan and Sadeghi, Ahmad-Reza},
booktitle = {USENIX Security Symposium (USENIX Security)},
year={2025}
}

@misc{Mitre,
	author = {{MITRE}},
	title = {Common Weakness Enumeration},
	note = {Last accessed: 11/18/2023}, 
	url = {https://cwe.mitre.org/data/index.html},
}

@misc{Morlkx,
	author = {OpenRISC},
	title = {mor1kx - an OpenRISC processor {IP} core},
	note = {Last accessed: 11/18/2023},
	url = {https://github.com/openrisc/mor1kx},
}

@misc{Verilator,
	author={Verilator}, 
	title={Welcome to Verilator},
	url={https://www.veripool.org/verilator/},
	note = {Last accessed: 11/18/2023},
}

@inproceedings{revizor-oleksenko,
author = {Oleksenko, Oleksii and Fetzer, Christof and K\"{o}pf, Boris and Silberstein, Mark},
title = {Revizor: testing black-box CPUs against speculation contracts},
year = {2022},
isbn = {9781450392051},
publisher = {Association for Computing Machinery},
address = {New York, NY, USA},
url = {https://doi.org/10.1145/3503222.3507729},
doi = {10.1145/3503222.3507729},
booktitle = {Proceedings of the 27th ACM International Conference on Architectural Support for Programming Languages and Operating Systems},
pages = {226–239},
numpages = {14},
location = {Lausanne, Switzerland},
series = {ASPLOS '22}
}

@inproceedings {medusa-moghimi,
author = {Daniel Moghimi and Moritz Lipp and Berk Sunar and Michael Schwarz},
title = {Medusa: Microarchitectural Data Leakage via Automated Attack Synthesis},
booktitle = {29th USENIX Security Symposium (USENIX Security 20)},
year = {2020},
isbn = {978-1-939133-17-5},
pages = {1427--1444},
url = {https://www.usenix.org/conference/usenixsecurity20/presentation/moghimi-medusa},
publisher = {USENIX Association},
month = aug
}

@inproceedings{absynthe-gras,
  title={ABSynthe: Automatic Blackbox Side-channel Synthesis on Commodity Microarchitectures.},
  author={Gras, Ben and Giuffrida, Cristiano and Kurth, Michael and Bos, Herbert and Razavi, Kaveh},
  booktitle={NDSS},
  year={2020}
}

@inproceedings{speechminer-xiao,
  author = {Xiao, Yuan and Zhang, Yinqian and Teodorescu, Radu},
  booktitle = {NDSS},
  ee = {https://www.ndss-symposium.org/ndss-paper/speechminer-a-framework-for-investigating-and-measuring-speculative-execution-vulnerabilities/},
  isbn = {1-891562-61-4},
  publisher = {The Internet Society},
  title = {SPEECHMINER: A Framework for Investigating and Measuring Speculative Execution Vulnerabilities.},
  url = {http://dblp.uni-trier.de/db/conf/ndss/ndss2020.html#XiaoZT20},
  year = 2020
}

@misc{osiris-weber,
      title={Osiris: Automated Discovery of Microarchitectural Side Channels}, 
      author={Daniel Weber and Ahmad Ibrahim and Hamed Nemati and Michael Schwarz and Christian Rossow},
      year={2021},
      eprint={2106.03470},
      archivePrefix={arXiv},
      primaryClass={cs.CR},
      url={https://arxiv.org/abs/2106.03470}, 
}

@inproceedings{observationrefinement, author = {Buiras, Pablo and Nemati, Hamed and Lindner, Andreas and Guanciale, Roberto}, title = {Validation of Side-Channel Models via Observation Refinement}, year = {2021}, isbn = {9781450385572}, publisher = {Association for Computing Machinery}, address = {New York, NY, USA}, url = {https://doi.org/10.1145/3466752.3480130}, doi = {10.1145/3466752.3480130}, booktitle = {MICRO-54: 54th Annual IEEE/ACM International Symposium on Microarchitecture}, pages = {578–591}, numpages = {14}, location = {Virtual Event, Greece}, series = {MICRO '21} }

@misc{scamv,
      title={Validation of Abstract Side-Channel Models for Computer Architectures}, 
      author={Hamed Nemati and Pablo Buiras and Andreas Lindner and Roberto Guanciale and Swen Jacobs},
      year={2020},
      eprint={2005.05254},
      archivePrefix={arXiv},
      primaryClass={cs.CR},
      url={https://arxiv.org/abs/2005.05254}, 
}

@inproceedings {cascade-razavi,
author = {Flavien Solt and Katharina Ceesay-Seitz and Kaveh Razavi},
title = {Cascade: {CPU} Fuzzing via Intricate Program Generation},
booktitle = {33rd USENIX Security Symposium (USENIX Security 24)},
year = {2024},
isbn = {978-1-939133-44-1},
address = {Philadelphia, PA},
pages = {5341--5358},
url = {https://www.usenix.org/conference/usenixsecurity24/presentation/solt},
publisher = {USENIX Association},
month = aug
}

@INPROCEEDINGS{feedbackfuzz,
  author={Wang, Jiashun and Cui, Baojiang and Dong, Renhai and Zhai, Rundi},
  booktitle={ICASSP 2025 - 2025 IEEE International Conference on Acoustics, Speech and Signal Processing (ICASSP)}, 
  title={FeedbackFuzz: Fuzzing Processors via Intricate Program Generation with Feedback Engine}, 
  year={2025},
  volume={},
  number={},
  pages={1-5},
  doi={10.1109/ICASSP49660.2025.10889404}}

@article{liblisa, author = {Craaijo, Jos and Verbeek, Freek and Ravindran, Binoy}, title = {libLISA: Instruction Discovery and Analysis on x86-64}, year = {2024}, issue_date = {October 2024}, publisher = {Association for Computing Machinery}, address = {New York, NY, USA}, volume = {8}, number = {OOPSLA2}, url = {https://doi.org/10.1145/3689723}, doi = {10.1145/3689723}, journal = {Proc. ACM Program. Lang.}, month = oct, articleno = {283}, numpages = {29}  }

@inproceedings{idev,
author = {Qin, Shisong and Zhang, Chao and Chen, Kaixiang and Li, Zheming},
title = {iDEV: exploring and exploiting semantic deviations in ARM instruction processing},
year = {2021},
isbn = {9781450384599},
publisher = {Association for Computing Machinery},
address = {New York, NY, USA},
url = {https://doi.org/10.1145/3460319.3464842},
doi = {10.1145/3460319.3464842},
booktitle = {Proceedings of the 30th ACM SIGSOFT International Symposium on Software Testing and Analysis},
pages = {580–592},
numpages = {13},
location = {Virtual, Denmark},
series = {ISSTA 2021}
}

@INPROCEEDINGS{iscanu,
  author={Dofferhoff, Rens and Göebel, Michael and Rietveld, Kristian and van der Kouwe, Erik},
  booktitle={2020 50th Annual IEEE/IFIP International Conference on Dependable Systems and Networks (DSN)}, 
  title={iScanU: A Portable Scanner for Undocumented Instructions on RISC Processors}, 
  year={2020},
  volume={},
  number={},
  pages={306-317},
  doi={10.1109/DSN48063.2020.00047}}

@misc{riscvuzz,
  title={{RISCVuzz}: Discovering Architectural {CPU} Vulnerabilities via Differential Hardware Fuzzing},
  author={Thomas, Fabian and Hetterich, Lorenz and Zhang, Ruiyi and Weber, Daniel and Gerlach, Lukas and Schwarz, Michael},
  howpublished = {\url{https://ghostwriteattack.com/}},
  year={2024},
}

@misc{domas-sandsifter,
	author = {Christopher Domas},
	title = {Breaking the x86 ISA},
	howpublished = {\url{https://www.blackhat.com/docs/us-17/thursday/us-17-Domas-Breaking-The-x86-Instruction-Set-wp.pdf}},
	year = {},
	note = {[Accessed 19-04-2025]},
}

@ARTICLE{uisfuzz,
  author={Li, Xixing and Wu, Zehui and Wei, Qiang and Wu, Haolan},
  journal={IEEE Access}, 
  title={UISFuzz: An Efficient Fuzzing Method for CPU Undocumented Instruction Searching}, 
  year={2019},
  volume={7},
  number={},
  pages={149224-149236},
  doi={10.1109/ACCESS.2019.2946444}}

@INPROCEEDINGS{mishegos,
  author={Woodruff, William and Carroll, Niki and Peters, Sebastiaan},
  booktitle={2021 IEEE Security and Privacy Workshops (SPW)}, 
  title={Differential analysis of x86-64 instruction decoders}, 
  year={2021},
  volume={},
  number={},
  pages={152-161},
  doi={10.1109/SPW53761.2021.00029}}

@inproceedings{n-version-disassambly, author = {Paleari, Roberto and Martignoni, Lorenzo and Fresi Roglia, Giampaolo and Bruschi, Danilo}, title = {N-version disassembly: differential testing of x86 disassemblers}, year = {2010}, isbn = {9781605588230}, publisher = {Association for Computing Machinery}, address = {New York, NY, USA}, url = {https://doi.org/10.1145/1831708.1831741}, doi = {10.1145/1831708.1831741}, booktitle = {Proceedings of the 19th International Symposium on Software Testing and Analysis}, pages = {265–274}, numpages = {10}, location = {Trento, Italy}, series = {ISSTA '10} }

@ARTICLE{skipscan,
  author={Wang, Guang and Zhu, Ziyuan and Cheng, Xu and Meng, Dan},
  journal={IEEE Transactions on Computers}, 
  title={A High-Coverage and Efficient Instruction-Level Testing Approach for x86 Processors}, 
  year={2023},
  volume={72},
  number={11},
  pages={3203-3217},
  doi={10.1109/TC.2023.3288762}}

@misc{examiner,
      title={Automatically Locating ARM Instructions Deviation between Real Devices and CPU Emulators}, 
      author={Muhui Jiang and Tianyi Xu and Yajin Zhou and Yufeng Hu and Ming Zhong and Lei Wu and Xiapu Luo and Kui Ren},
      year={2021},
      eprint={2105.14273},
      archivePrefix={arXiv},
      primaryClass={cs.CR},
      url={https://arxiv.org/abs/2105.14273}, 
}

@misc{encarsia,
	author = {Bölcskei, Matej and Solt, Flavien  and Ceesay-Seitz, Katharina and Razavi, Kaveh},
	title = {{Encarsia}: Evaluating CPU Fuzzers via Automatic Bug Injection},
	howpublished = {\url{https://comsec.ethz.ch/wp-content/files/encarsia_sec25.pdf}},
	year = {2025},
	note = {[Accessed 19-04-2025]},
}

@inproceedings {feedback-in-fuzzing,
author = {Jinghan Wang and Yue Duan and Wei Song and Heng Yin and Chengyu Song},
title = {Be Sensitive and Collaborative: Analyzing Impact of Coverage Metrics in Greybox Fuzzing},
booktitle = {22nd International Symposium on Research in Attacks, Intrusions and Defenses (RAID 2019)},
year = {2019},
isbn = {978-1-939133-07-6},
address = {Chaoyang District, Beijing},
pages = {1--15},
url = {https://www.usenix.org/conference/raid2019/presentation/wang},
publisher = {USENIX Association},
month = sep
}

@manual{x86-64,
 title = "{I}ntel 64 and {IA}-32 {A}rchitectures {S}oftware {D}eveloper’s {M}anual",
}

@techreport{boom,
    Author= {Celio, Christopher and Patterson, David A. and Asanović, Krste},
    Title= {The Berkeley Out-of-Order Machine (BOOM): An Industry-Competitive, Synthesizable, Parameterized RISC-V Processor},
    Year= {2015},
    Month= {Jun},
    Url= {http://www2.eecs.berkeley.edu/Pubs/TechRpts/2015/EECS-2015-167.html},
    Number= {UCB/EECS-2015-167},
}

@misc{chipsandcheeseLionCove,
	author = {Chester Lam},
	title = {{L}ion {C}ove: {I}ntel’s {P}-{C}ore {R}oars --- chipsandcheese.com},
	howpublished = {\url{https://chipsandcheese.com/p/lion-cove-intels-p-core-roars}},
	year = {},
	note = {[Accessed 24-04-2025]},
}

@misc{verificationacademyCoverage,
	author = {Foster, Harry},
	title = {{C}overage --- verificationacademy.com},
	howpublished = {\url{https://verificationacademy.com/topics/coverage/}},
	year = {},
	note = {[Accessed 24-04-2025]},
}

@inproceedings{logicfuzzer-jose-renau,
author = {Kabylkas, Nursultan and Thorn, Tommy and Srinath, Shreesha and Xekalakis, Polychronis and Renau, Jose},
title = {Effective Processor Verification with Logic Fuzzer Enhanced Co-simulation},
year = {2021},
isbn = {9781450385572},
publisher = {Association for Computing Machinery},
address = {New York, NY, USA},
url = {https://doi.org/10.1145/3466752.3480092},
doi = {10.1145/3466752.3480092},
booktitle = {MICRO-54: 54th Annual IEEE/ACM International Symposium on Microarchitecture},
pages = {667–678},
numpages = {12},
location = {Virtual Event, Greece},
series = {MICRO '21}
}

@misc{riscv-dv,
	author = {{taoliug} and Sharma, Anil and Goel, Puneet and Esfeden, Hodjat Asghari and Dang, Hai Hoang and {weicaiyang} and van der Maas, Marno and Patel, Yash and Kurc, Maciej and Chadwick, Greg and Johnson, Scott and Devaiya, Shraddha and {danielmlynek} and {ishita71} and Stacha{\' n}czyk, Dariusz and Moore, Lee and {simond-imperas} and Michalak, Tomasz and {MateuszKarlic} and Fegran, Henrik and Gugala, Karol and Kaliraj, Pradheep and {wkkuna} and Swarbrick, Rupert and {shrujal20} and Singh, Saurabh and Vogel, Pirmin and Wagner, Philipp and Callahan, Harry and Topal, Canberk},
	year = {2025},
	month = {jun 5},
	title = {chipsalliance/riscv-dv},
	url = {https://github.com/chipsalliance/riscv-dv},
	howpublished = {https://github.com/chipsalliance/riscv-dv},
}

@misc{riscv-tests,
	author = {{RISC-V Foundation}},
	title = {{RISCV-Tests: riscv-software-src/riscv-tests}},
	howpublished = {\url{https://github.com/riscv-software-src/riscv-tests}},
	year = {},
	note = {[Accessed 12-08-2025]},
}

@article{fuzzing-seminal-unix-utilities,
author = {Miller, Barton P. and Fredriksen, Lars and So, Bryan},
title = {An empirical study of the reliability of UNIX utilities},
year = {1990},
issue_date = {Dec. 1990},
publisher = {Association for Computing Machinery},
address = {New York, NY, USA},
volume = {33},
number = {12},
issn = {0001-0782},
url = {https://doi.org/10.1145/96267.96279},
doi = {10.1145/96267.96279},
journal = {Commun. ACM},
month = dec,
pages = {32–44},
numpages = {13}
}

@misc{covert-shotgun,
	author = {Anders Fogh},
	title = {""{C}overt {S}hotgun"},
	howpublished = {\url{https://cyber.wtf/2016/09/27/covert-shotgun/}},
	year = {September 2016},
	note = {[Accessed 25-08-2025]},
}

@inproceedings{amulet-gururaj,
author = {Fu, Bo and Tenenbaum, Leo and Adler, David and Klein, Assaf and Gogia, Arpit and Alameldeen, Alaa R. and Guarnieri, Marco and Silberstein, Mark and Oleksenko, Oleksii and Saileshwar, Gururaj},
title = {AMuLeT: Automated Design-Time Testing of Secure Speculation Countermeasures},
year = {2025},
isbn = {9798400710797},
publisher = {Association for Computing Machinery},
address = {New York, NY, USA},
url = {https://doi.org/10.1145/3676641.3716247},
doi = {10.1145/3676641.3716247},
booktitle = {Proceedings of the 30th ACM International Conference on Architectural Support for Programming Languages and Operating Systems, Volume 2},
pages = {32–47},
numpages = {16},
location = {Rotterdam, Netherlands},
series = {ASPLOS '25}
}

@ARTICLE{lstm-schmidhuber,
  author={Hochreiter, Sepp and Schmidhuber, Jürgen},
  journal={Neural Computation}, 
  title={Long Short-Term Memory}, 
  year={1997},
  volume={9},
  number={8},
  pages={1735-1780},
  doi={10.1162/neco.1997.9.8.1735}}

@ARTICLE{rl-introduction-sutton,
  author={Sutton, R.S. and Barto, A.G.},
  journal={IEEE Transactions on Neural Networks}, 
  title={Reinforcement Learning: An Introduction}, 
  year={1998},
  volume={9},
  number={5},
  pages={1054-1054},
  doi={10.1109/TNN.1998.712192}}

@inproceedings{gpt-radford,
  title={Improving Language Understanding by Generative Pre-Training},
  author={Alec Radford and Karthik Narasimhan},
  year={2018},
  url={https://cdn.openai.com/research-covers/language-unsupervised/language_understanding_paper.pdf}
}

@misc{iverilog,
	author = {},
	title = {{G}it{H}ub - steveicarus/iverilog: {I}carus {V}erilog --- github.com},
	howpublished = {\url{https://github.com/steveicarus/iverilog}},
	year = {},
	note = {[Accessed 17-11-2025]},
}

@inproceedings{saravanan2024Odyssey,
author = {Saravanan, Raghul and Pudukotai Dinakarrao, Sai Manoj},
title = {The Fuzz Odyssey: A Survey on Hardware Fuzzing Frameworks for Hardware Design Verification},
year = {2024},
isbn = {9798400706059},
publisher = {Association for Computing Machinery},
address = {New York, NY, USA},
url = {https://doi.org/10.1145/3649476.3658697},
doi = {10.1145/3649476.3658697},
booktitle = {Proceedings of the Great Lakes Symposium on VLSI 2024},
pages = {192–197},
numpages = {6},
location = {Clearwater, FL, USA},
series = {GLSVLSI '24}
}

@misc{nathCADforAssurance,
	author = {Atul Prasad Deb Nath},
	title = {SoC Benchmarks-CAD for Assurance},
	note = {Last accessed: 11/18/2025},
	url = {https://cadforassurance.org/soc-platform/soc-benign-benchmark/system-on-chip-benchmarks/},
}

@inproceedings{metzman2021FuzzBench, author = {Metzman, Jonathan and Szekeres, L\'{a}szl\'{o} and Simon, Laurent and Sprabery, Read and Arya, Abhishek}, title = {FuzzBench: an open fuzzer benchmarking platform and service}, year = {2021}, isbn = {9781450385626}, publisher = {Association for Computing Machinery}, address = {New York, NY, USA}, url = {https://doi.org/10.1145/3468264.3473932}, doi = {10.1145/3468264.3473932}, booktitle = {Proceedings of the 29th ACM Joint Meeting on European Software Engineering Conference and Symposium on the Foundations of Software Engineering}, pages = {1393–1403}, numpages = {11}, location = {Athens, Greece}, series = {ESEC/FSE 2021} }

@misc{synopsys-vcs,
	author = {},
	title = {{V}{C}{S}: {F}unctional {V}erification {S}olution | {S}ynopsys --- synopsys.com},
	howpublished = {\url{https://www.synopsys.com/verification/simulation/vcs.html}},
	year = {},
	note = {[Accessed 27-08-2025]},
}

@misc{cadence-xcelium,
	author = {},
	title = {{X}celium {L}ogic {S}imulation --- cadence.com},
	howpublished = {\url{https://www.cadence.com/en_US/home/tools/system-design-and-verification/simulation-and-testbench-verification/xcelium-simulator.html}},
	year = {},
	note = {[Accessed 27-08-2025]},
}

@misc{siemens-questa,
	author = {},
	title = {{Q}uesta {O}ne {S}im --- eda.sw.siemens.com},
	howpublished = {\url{https://eda.sw.siemens.com/en-US/ic/questa-one/simulation/questa-one-sim/}},
	year = {},
	note = {[Accessed 27-08-2025]},
}
%\appendix

\appendices
\section*{Appendix}
\renewcommand{\thesection}{\Alph{section}}

% In Encarsia~\cite{encarsia}, a tool designed to inject synthetic bugs in designs to evaluate RTL fuzzers, the authors survey four RTL fuzzing works (Cascade, DiFuzzRTL, ProcessorFuzz, and RFUZZ) and identify eight insights and provide five suggestions to guide future fuzzing. Our work not only echoes the same insights by analyzing a broader body of work, but also identifies more insights across different abstraction levels and discusses how to address those concerns/insights through future work. Additionally, we systematize the existing body of work through our classification framework and distill the current state of the field though this paper.

\section{SW Fuzzing to HW Fuzzing}  \label{sec:appendix-SW-to-HW}

The general principles of software fuzzing extend to hardware fuzzing with specific adaptations. Software fuzzing typically comprises three steps: (1) input generation and pre-processing, (2) evaluation, and (3) feedback-driven refinement of subsequent inputs. Hardware fuzzing follows the same overall pattern, but each step is adapted according to the target type (black-box, grey-box, or white-box), the class of bugs being targeted, and hardware-specific considerations.

\textbf{Input generation:} In hardware fuzzing, inputs can be generated using model-less or model-based approaches, or through mutation-based techniques. Hardware designs generally include explicit specifications for expected input formats, and deviations from these specifications can result in undefined behavior, such as processor freezes or resets. In black-box CPU fuzzing (ISA fuzzing and microarchitectural fuzzing in this paper), inputs typically consist of assembly instructions, which must comply with ISA-specific encodings and semantics to be considered valid. These encodings act as a simple input model or grammar. Randomized (model-less) input generation can also be employed, though with limited effectiveness, as demonstrated in this paper. Input mutation is commonly used as well; however, unless the objective is to explore undocumented behaviors or instructions, mutations generally adhere to the target’s input specification or incorporate goal-specific modifications derived from prior knowledge of the target.
The general steps and ideas of software fuzzing are carried over to hardware fuzzing with some adaptations. Software fuzzing typically contains the following steps: 1) input generation and pre-processing, 2) evaluation, and 3) feedback-driven refinement of future inputs. Hardware fuzzing follows the same pattern but each step is adapted based on the target (black-box, grey-box, or white-box), type of bugs targeted in addition to hardware-specific adaptations.

\textbf{Pre-processing/Instrumentation:} In software fuzzing, pre-processing and instrumentation typically involve the use of sanitizers and profiling tools to instrument code for data collection without affecting functionality. In hardware fuzzing, instrumentation is generally feasible only in the context of RTL fuzzing, where the RTL design is modified to include custom coverage points, information flow tracking, memory hashing logic, and hardware error handlers. It should be noted that such RTL instrumentation can slow simulations or emulations by several orders of magnitude. For ISA fuzzing, pre-processing entails identifying the device tree to implement custom exception handlers and creating wrapper programs to execute test programs on the target while collecting execution traces. In some cases, ISA emulators, simulators, or real CPUs are used to validate generated inputs as part of this step. In microarchitecture (uArch) fuzzing, pre-processing involves devising side channels (e.g., dcache-based flush–reload channels) to expose bugs and provide coverage information that guides fuzzing.

\textbf{Evaluation:} Similar to software fuzzing, the evaluation step in hardware fuzzing involves executing the generated test cases on the target. However, unlike software, where an out-of-bounds memory access typically results in a segmentation fault, in hardware fuzzing with bare-metal test programs (i.e., no operating system), out-of-bounds accesses remain undetected as long as they stay within the processor’s data memory region. Conversely, accesses to kernel pages or I/O memory regions trigger exceptions, which cause the test program to fail if not properly handled. The logs produced by the test harness often contain limited information, requiring manual or software-assisted analysis to identify the offending instruction. For instance, an x86 core raises a \#UD exception upon encountering an undefined instruction, necessitating post hoc investigation to determine why the instruction was considered illegal by the hardware.

It should be noted that a buggy processor may raise exceptions even on valid instructions. For this reason, hardware fuzzers heavily rely on a Golden Reference Model (GRM). A GRM can take various forms, including an ISA simulator, emulator, disassembler, another CPU, an abstract model of the CPU, or an abstract model of a submodule within the CPU. The differential fuzzing (DF), observation-independent fuzzing (OIF), and model-based relational testing (MRT) methods studied in this paper, though distinct in methodology, all execute the generated inputs on both the target and the GRM. Any deviation between the target and the GRM flags the input for further analysis to identify bugs in either the target or the reference model. In MRT approaches, abstract and observation models often augment an emulator with security properties or formally define the expected behavior of a submodule.

\textbf{Feedback:} The final step in software fuzzing is the feedback loop, which refines subsequent test inputs. In grey-box or white-box hardware fuzzing, where coverage information can be extracted, similar feedback loops are employed. However, for black-box hardware targets with limited or no observability, obtaining reliable feedback is difficult and often omitted. White-box and grey-box hardware fuzzers that incorporate feedback typically adopt AFL’s mutation strategies to guide input generation.

\section{AFL Mutations used in Hardware Fuzzing}\label{sec:afl}

The majority of hardware fuzzers utilize American Fuzzy Lop (AFL) mutation operators to generate new inputs from seed programs. Table~\ref{tab:hw-sw-appendix-afl-mutations} outlines several AFL mutation operators commonly employed in hardware fuzzing. These operators can be applied to various types of input seeds, including sequences of bits as well as RISC-V or x86 instructions. 

\begin{table}[!htb]
\resizebox{\linewidth}{!}{
    \begin{tabular}{|l|p{5cm}|}
    \hline  
      \textbf{Name} & \textbf{Description}  \\
      \hline

    Bitflip 1/1 &  Flip single bit \\
     \hline 
    Bitflip 2/1 &   Flip two adjacent bits\\
     \hline
    Bitflip 4/1 & Flip four adjacent bits \\
     \hline
    Bitflip 8/8 &  Flip single byte \\
     \hline
    Bitflip 32/8 &  Flip four adjacent bytes \\
     \hline
    Arith 8/8  &  Treat single byte as 8-bit integer, add/sub
values from 0 to 35 \\
     \hline
    Interest 8  &  Overwrite a random 8-bit integer with interesting
value \\
     \hline
    Interest 16 &   Overwrite a random 16-bit integer with interesting
value \\
     \hline
    Interest 32 &  Overwrite a random 32-bit integer with interesting
value \\
     \hline
    Random 8 &  Overwrite random byte with random value \\
     \hline
       Delete &  Delete a random sequence of bytes \\
     \hline
       Clone & Clone a random sequence of bytes \\
     \hline
    
    \end{tabular}
    }
    \caption{ {AFL Mutations used by HW Fuzzers} }
    \label{tab:hw-sw-appendix-afl-mutations}
\end{table}

\section{Comparison of Coverage Metrics} \label{tab:appendix-coverage-metrics}

\begin{table}[h]
\centering
\resizebox{\linewidth}{!}{%
\begin{tabular}{|p{1.4cm}|p{2.5cm}|p{2.5cm}|p{2.5cm}|}
\hline
\textbf{Coverage Type} & \textbf{Definition} & \textbf{Use Case} & \textbf{Limitation} \\ \hline
Branch        & Outcomes of conditional statements & Control flow  coverage   & Poor scalability in large designs \\ \hline
Toggle        & 0$\rightarrow$1/1$\rightarrow$0 transitions of signals & Broad observability        & Large number of signals; poor coverage metric \\ \hline
Finite State Machines (FSM)          & State transitions in FSMs           & Identify Control logic, rare states in design  & High instrumentation overhead \\ \hline
Expression/ Line & Evaluation of every HDL line & Debugging and fine-grained RTL verification & Costly, may miss deeper bugs \\ \hline
Functional    & User-defined functionality points   & Specification compliance             & Manual effort and human bias \\ \hline
Multiplexer (MUX)      & Toggling of MUX select signals      & Datapath exploration        & Nonscalable and clock-insensitive \\ \hline
Register      & Register toggling & Scalable and low overhead      & Misses high-level bugs \\ \hline
CSR Transition & Transitions of Control \& Status Registers & No HDL instrumentation required and temporal bugs & Limited scope and false positives \\ \hline
\end{tabular}%
}
\caption{Comparison of coverage metrics: definitions, use cases, and trade-offs.}
\label{tab:coverage-metrics}

\end{table}

Table~\ref{tab:coverage-metrics} serves as an initial reference to help users understand common coverage metrics, their use cases, and associated limitations. Commercial EDA tools typically combine Branch, FSM, Expression, and Line coverage metrics, which incur substantial instrumentation overhead, as discussed in Section~\ref{sec:typec-coverage-disc}. However, these tools can also be configured to measure each metric independently.

\section{Selecting a Hardware Fuzzer} \label{sec:appendix-select-fuzzer}

\begin{sidewaysfigure*}
    \centering
    \includegraphics[width=\linewidth]{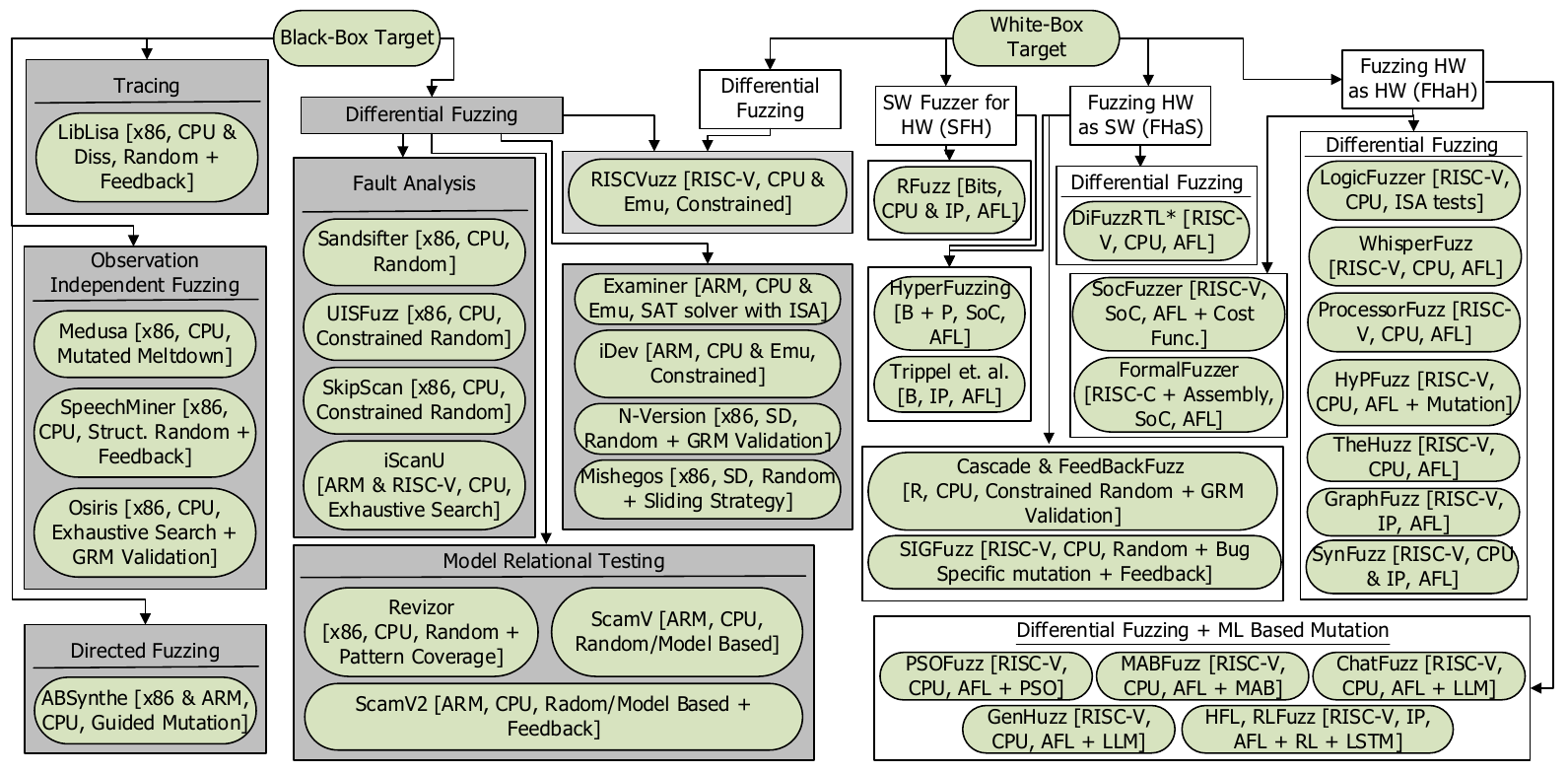}
    \caption{Flowchart to select a hardware fuzzer. This chart is to be used in conjunction with Table~\ref{tab:comparison}. * - DiFuzzRTL also has a FPGA simulation feature that could provide the advantages of FHaH methodology.}
    \label{fig:appendix-flowchart}
\end{sidewaysfigure*}

As discussed in the paper, selecting an appropriate hardware fuzzer depends on several factors: (1) the type of target (black-box or white-box), (2) the class of bugs to be identified (e.g., functional bugs, side channels, undisclosed instructions), and (3) the preferred fuzzing methodology (e.g., tracing, differential fuzzing, directed fuzzing, model-based relational testing). Flowchart~\ref{fig:appendix-flowchart} presents a multi-step process for selecting a fuzzer based on the target type and the chosen methodology.

For instance, if a user wishes to verify a white-box target using differential fuzzing combined with ML-based mutation strategies, they will have six options to choose from. Each option is annotated with a subset of its characteristics, denoted as [Input Type, Target Type, Fuzzing Algorithm]. If the user intends to verify an IP core, the choices narrow to two fuzzers: HFL and RLFuzz. Conversely, if the goal is to verify a CPU using an LLM-based approach, the available options are GenHuzz and ChatFuzz. To further differentiate between these options, the user may consult Table~\ref{tab:comparison} and the detailed analysis provided in the main body of the paper.

\begin{comment}
\section{Data Collection}
Add information about how we collect data for our comparisons. Point out situations where we are comparing data from two different papers. Point out why or why not such comparison is valid/invalid.

\section{Notes for Resubmission}
1. Insight numbers \\
2. highlight definitions for each section (SFH, FHaS, FHaH) \\
3. Highlight why each abstraction is different and also highlight works can span two abstraction levels. \\
4. Fuzz Odyssey \\
5. Restructure to make it clear that ISA and uArch are silicon-black boxes and RTL is pre-silicon white-box. Reviewer comment "If the classification is about whether the inputs are provided to a silicon implementation or not, then is the ISA class about architectural bugs in silicon, uarch about timing side channels in silicon, and RTL can be like ISA or uarch, but pre-silicon (e.g., TheHuzz looks for architectural bugs and WhisperFuzz looks for timing side channels)?" \\
6. Future Work - Piggyback on existing cross-layer work. Tell it has potential, We can do better. More research in that area. \\
\end{comment}

\end{document}